\documentclass[twocolumn]{aastex63}
\usepackage{amsmath,color,textcomp,url,graphicx}

\usepackage{subfigure}
\makeatletter

\renewcommand{\@thesubfigure}{(\alph{subfigure})\hskip\subfiglabelskip}
\renewcommand{\@@thesubfigure}{(\alph{subfigure})}
\makeatother

\newcommand{\kms}{\hbox{km\,s$^{-1}$}}
\newcommand{\ms}{\hbox{m\,s$^{-1}$}}
\newcommand{\masyr}{$\mathrm{mas}\,\mathrm{yr}^{-1}$}
\newcommand{\age}{$510 \pm 95$}
\newcommand{\totalmembers}{50}
\newcommand{\shortage}{$\approx 500$}
\newcommand{\msun}{$M_{\mathrm{Sun}}$}

\newcommand{\mjup}{$M_{\mathrm{Jup}}$}

\newcommand{\logrhk}{$\log R^\prime_{\rm HK}$}

\received{2022 July 29}
\revised{2022 October 26}

\submitjournal{ApJ}

\shorttitle{The Oceanus Moving Group}
\shortauthors{Gagn\'e et al.}

\begin{document}

\title{THE OCEANUS MOVING GROUP: A NEW 500 Myr-OLD HOST FOR THE NEAREST BROWN DWARF}

\author[0000-0002-2592-9612]{Jonathan Gagn\'e}
\affiliation{Plan\'etarium Rio Tinto Alcan, Espace pour la Vie, 4801 av. Pierre-de Coubertin, Montr\'eal, Qu\'ebec, Canada}
\affiliation{Institute for Research on Exoplanets, Universit\'e de Montr\'eal, D\'epartement de Physique, C.P.~6128 Succ. Centre-ville, Montr\'eal, QC H3C~3J7, Canada}
\email{gagne@astro.umontreal.ca}
\author[0000-0001-7171-5538]{Leslie Moranta}
\affiliation{Plan\'etarium Rio Tinto Alcan, Espace pour la Vie, 4801 av. Pierre-de Coubertin, Montr\'eal, Qu\'ebec, Canada}
\affiliation{Institute for Research on Exoplanets, Universit\'e de Montr\'eal, D\'epartement de Physique, C.P.~6128 Succ. Centre-ville, Montr\'eal, QC H3C~3J7, Canada}
\author[0000-0001-6251-0573]{Jacqueline K. Faherty}
\affiliation{Department of Astrophysics, American Museum of Natural History, Central Park West at 79th St., New York, NY 10024, USA}
\author[0000-0003-2102-3159]{Rocio Kiman}
\affil{Kavli Institute for Theoretical Physics, University of California, Santa Barbara, CA 93106, USA}
\author[0000-0003-2604-3255]{Dominic Couture}
\affiliation{Plan\'etarium Rio Tinto Alcan, Espace pour la Vie, 4801 av. Pierre-de Coubertin, Montr\'eal, Qu\'ebec, Canada}
\affiliation{Institute for Research on Exoplanets, Universit\'e de Montr\'eal, D\'epartement de Physique, C.P.~6128 Succ. Centre-ville, Montr\'eal, QC H3C~3J7, Canada}
\author[0000-0002-3499-2998]{Arnaud Ren\'e Larochelle}
\affiliation{Plan\'etarium Rio Tinto Alcan, Espace pour la Vie, 4801 av. Pierre-de Coubertin, Montr\'eal, Qu\'ebec, Canada}
\affiliation{Institute for Research on Exoplanets, Universit\'e de Montr\'eal, D\'epartement de Physique, C.P.~6128 Succ. Centre-ville, Montr\'eal, QC H3C~3J7, Canada}
\author[0000-0001-9482-7794]{Mark Popinchalk}
\affil{Department of Astrophysics, American Museum of Natural History, Central Park West at 79th St., New York, NY 10024, USA}
\affiliation{Physics, The Graduate Center, City University of New York, New York, NY 10016, USA}
\affiliation{Department of Physics and Astronomy, Hunter College, City University of New York, 695 Park Avenue, New York, NY 10065, USA}
\author[0000-0002-1169-1834]{Daniella Morrone}
\affiliation{Plan\'etarium Rio Tinto Alcan, Espace pour la Vie, 4801 av. Pierre-de Coubertin, Montr\'eal, Qu\'ebec, Canada}
\affiliation{Institute for Research on Exoplanets, Universit\'e de Montr\'eal, D\'epartement de Physique, C.P.~6128 Succ. Centre-ville, Montr\'eal, QC H3C~3J7, Canada}

\begin{abstract}

We report the discovery of the Oceanus moving group, a \shortage\,Myr-old group with \totalmembers\ members and candidate members at distances 2--50\,pc from the Sun using an unsupervised clustering analysis of nearby stars with Gaia~DR3 data. This new moving group includes the nearest brown dwarf WISE~J104915.57--531906.1~AB (Luhman~16~AB) at a distance of 2\,pc, which was previously suspected to be young ($600-800$\,Myr) based on a comparison of its dynamical mass measurements with brown dwarf evolutionary models. We use empirical color-magnitude sequences, stellar activity and gyrochronology to determine that this new group is roughly coeval with the Coma Ber open cluster, with an isochronal age of $510 \pm 95$\,Myr. This newly discovered group will be useful to refine the age and chemical composition of Luhman~16~AB, which is already one of the best substellar benchmarks known to date. Furthermore, the Oceanus moving group is one of the nearest young moving groups identified to date, making it a valuable laboratory for the study of exoplanets and substellar members, with 8 brown dwarf candidate members already identified here.

\end{abstract}


\section{INTRODUCTION}\label{sec:intro}

Star formation from a molecular cloud gives rise to ensembles of typically a few hundred stars in a short period that share the same galactic orbit as their parent cloud \citep{2003ARAA..41...57L}. As they complete a few orbits around the Galaxy, these loose moving groups tend to become disrupted by the irregular galactic potential and gradually migrate to distinct orbits, making it much harder to assign a star to its parent moving group after a few hundred million years. Spatially denser open clusters, which can contain thousands of stars, take longer to become entirely disrupted and we can therefore recognize their co-moving (and therefore coeval) nature for longer \citep{2003MNRAS.338..717B}.

The advent of the Gaia mission \citep{2016AA...595A...1G} sparked a revolution in our understanding and mapping of ensembles of young, coeval stars. Gaia provides accurate trigonometric distances and proper motions for more than a billion stars, and more recently with the third data release (DR3; \citealp{2021AA...649A...1G,gaiadr3}), it also provides heliocentric radial velocities for 33 million stars based on multi-epoch measurements thereby making it possible to calculate the 3D positions (the $XYZ$ Galactic coordinates) and 3D velocities (the $UVW$ space velocities) for a much larger ensemble of stars than was previously possible. This data set has rapidly allowed the scientific community to discover extended tidal tails around dense open clusters older than a few hundred million years (e.g., the Hyades, \citealp{2019AA...621L...2R}; Coma~Ber, \citealp{2019ApJ...877...12T}; and Praesepe, \citealp{2019AA...627A...4R}). In parallel, entirely new moving groups were discovered by the hundreds (e.g., see \citealp{2019AJ....158..122K}), some of which were found to be associated with star-forming regions (e.g., see \citealp{2019MNRAS.489.4418J,2021ApJS..254...20L}). In all, these new groups paint a complex picture of star formation that takes place in distinct stages along extended filaments.

The discovery and characterization of these coeval stellar populations is useful to identify benchmark objects with precisely known ages and chemical compositions, as it is difficult to measure these quantities precisely for a single star \citep{2010ARAA..48..581S}. A large collection of stars allows one to combine several different age-dating methods across a large number of stars, some of which can only be used with specific types of stars (e.g., lithium abundances, stellar activity through spectroscopic indices, UV or X-ray, gyrochronology, white dwarf cooling rates, or positions in a color-magnitude diagram). Some of these benchmarks have recently included substellar objects that were discovered as members of known moving groups, for which knowing their age is crucial to estimate their mass. This made it possible to recognize that some objects previously thought to be nearby brown dwarfs have masses below the Deuterium burning limit (e.g., see \citealt{2015ApJ...808L..20G}), meaning they would be categorized as exoplanets if they were orbiting a star. Such isolated planetary-mass objects defy current categories and are valuable to better understand the atmospheres of gas giant exoplanets given that we can obtain high-resolution, high-signal to noise spectra of their atmospheres, unhampered by the light of a much brighter star in their immediate vicinity (e.g., see \citealp{2013ApJ...777L..20L,2018ApJ...854L..27G}).

We report here the discovery of a new, slightly evolved moving group (\shortage\,Myr) which dissipation is likely well under way, in the immediate vicinity of the Sun (2--50\,pc). This new moving group is particularly interesting because it hosts the nearest brown dwarf system WISE~J104915.57--531906.1~AB (Luhman~16~AB, \citealp{2013ApJ...767L...1L}) that is already one of the best substellar benchmarks available to us because of its proximity ($1.9980 \pm 0.0004$\,pc; \citealp{2015MNRAS.453L.103S}) and its binary nature. Its component masses were measured through Kepler's laws ($27.9_{-1.0}^{+1.1}$\,\mjup\ and $34.2_{-1.1}^{+1.3}$\,\mjup; \citealp{2017ApJ...846...97G}); comparing these masses to substellar evolutionary models indicated that Luhman~16~AB appeared to be relatively young (600--800\,Myr), despite its kinematics not matching a known young association\footnote{A first analysis of Luhman~16~AB's motion hinted at a possible membership in the Argus association which suffers from a high rate of false-positive memberships. At 45--50\,Myr, the age of Argus appeared in tension with dynamical mass measurements, and further kinematics ruled out the possibility that Luhman~16~AB is related to Argus.}. The two components of Luhman~16~AB also have spectral types L7.5 and T0.5 \citep{2013ApJ...767L...1L}, making them among the rare brown dwarfs that straddle the so-called L/T transition where brown dwarf photospheres transition from cloudy to partially cloudy \citep{2008ApJ...689.1327S,2021ApJ...914..124B}, giving rise to a higher rate of photometric variability \citep{2020AJ....160...38V} and a sudden shift to bluer $J-K$ colors compared to warmer brown dwarfs in the L spectral class \citep{1999ApJ...522L..65B}. Doppler imaging of Luhman~16~B by \cite{2014Natur.505..654C} provided further evidence for the presence of patchy clouds on its surface in the form of large-scale bright and dark features on its surface. The recognition that Luhman~16~AB is a member of a young, nearby moving group opens the door to an independent age measurement for this system through a study of the other members in the moving group, and makes Luhman~16~AB an even more compelling substellar benchmark.

The new moving group identified here, which we named the Oceanus moving group, was found through an unsupervised clustering analysis directly in $XYZUVW$ space made possible by Gaia~DR3, as described in Section~\ref{sec:clustering}. We describe our search for additional members in Section~\ref{sec:addmem}, and a discussion of its individual members, population properties and age in Section~\ref{sec:disc}. We conclude this work in Section~\ref{sec:conclusion}.

\section{CLUSTERING ANALYSIS}\label{sec:clustering}

We used the HDBSCAN algorithm \citep{2017arXiv170507321M} to search for new clusters of stars within 200\,pc of the Sun based on Gaia~DR3 data with a method similar to that of \cite{2022arXiv220604567M}. HDBSCAN is a relatively fast algorithm that allows users to identify over-densities in $N$ dimensions in the presence of noisy data, and was recently used to identify several young stellar clusters (e.g., \citealp{2019AJ....158..122K,2021ApJ...917...23K}). The identification of young moving groups within about 100\,pc of the Sun has been challenging because most of these studies sacrificed the use of heliocentric radial velocities to analyze much larger sample sizes at larger distances, however, including full 6-dimensional kinematics is required to efficiently recover the nearest moving groups which tangential velocities span a large range of values and are strongly correlated with sky position (e.g., see \citealt{2022arXiv220604567M}). We therefore elected to use the full 6D kinematics provided by Gaia~DR3 to identify new, nearby candidate moving groups within 200\,pc of the Sun.

We found that the third Gaia data release provided a cleaner sample of stars near the Sun compared with Gaia~EDR3, which previously required a quality cut in parallax errors (which \citealp{2022arXiv220604567M} set at 0.4\,mas) to avoid including distant stars in the Galactic plane that appeared to have a large parallax due to cross-match errors. Furthermore, nearby Gaia~DR3 stars with full kinematics typically have 10--40 individual radial velocity epochs from which the heliocentric radial velocities are estimated (versus 2--20 for Gaia~DR2), making the space velocities much more robust against radial velocity variations caused by multiple systems. For these reasons, we have found that no quality cuts other than requiring a non-null radial velocity and a parallax above 5\,mas were required to obtain a good-quality sample to analyze with HDBSCAN.

Among other groups that will be presented in an upcoming publication (Moranta et al., in preparation), this analysis yielded the discovery of the nearby Oceanus moving group, which we named following the tradition of \cite{2019AJ....158..122K} where new groups discovered using Gaia data are named after children of the Gaia goddess from the Greek mythology. Here, we focus on the report and the preliminary characterization of the Oceanus moving group because of its potential at providing a robust age estimate for the nearest brown dwarf Luhman~16~AB. The initial list of Oceanus members uncovered by HDBSCAN contains 19 stars which are listed in Table~\ref{tab:candidates}.

\section{A SEARCH FOR ADDITIONAL MEMBERS}\label{sec:addmem}

We used the Bayesian model selection tool BANYAN~$\Sigma$ \citep{2018ApJ...856...23G} to identify additional members of the Oceanus moving group that were not initially included in our sample, most often because they lack full $XYZUVW$ kinematics. In order to do so, we built a single multivariate Gaussian model for the members described in Section~\ref{sec:clustering}, following a slightly updated version of the method described in \cite{2018ApJ...856...23G}. This method consists in building a 6-dimensional covariance matrix in $XYZUVW$ space for the list of Oceanus members, and then performing a Monte Carlo simulation as described in Section~7 of \cite{2018ApJ...856...23G} to assign a 82\% (for stars without radial velocities) or 90\% (for stars with a radial velocity) recovery rate when a 90\% probability threshold is used to select candidate members. Unlike the method of \cite{2018ApJ...856...23G} that did not fully account for measurement errors in the construction of the covariance matrix, we used the `extreme deconvolution' algorithm of \cite{2011AnApS...5.1657B} to properly account for the full error covariance matrix of each individual star. This allows us to obtain the intrinsic dispersion that is due to the physical spread in space and velocity of the Oceanus members, uncontaminated by the measurement errors. While the individual Oceanus members have a velocity dispersion of 0.73\,\kms\ in $UVW$ space, the resulting kinematic model is indicative of an intrinsic dispersion of 0.56\,\kms\ after excluding the impact of measurement errors. Young, coeval and co-moving stars usually have velocity dispersions in the range 0.3--1\,\kms, with which the dispersion of Oceanus members is consistent. We found average $XYZ$ Galactic coordinates and $UVW$ space velocities of $\left(11.45,12.69,15.98\right)$\,pc and $\left(-18.2,-24.3,-8.2\right)$\,\kms\ for the Oceanus moving group.

It must be noted that these average properties and the associated BANYAN~$\Sigma$ multivariate Gaussian model only includes the Oceanus members recovered automatically by HDBSCAN, as described in Section~\ref{sec:clustering}. The subsequent searches based on this model would therefore miss any spatial extension not initially detected by HDBSCAN using our sample based on Gaia~DR3, by construction.

\subsection{Missing Stellar Members}\label{sec:stellar}

We used the resulting spatial-kinematic model in combination with the other models already described in \cite{2018ApJ...856...23G} to identify additional members of the Oceanus moving group using both Gaia~DR3 and the 2007 re-reduction of the Hipparcos parallaxes \citep{2007AA...474..653V}, the latter of which we included to consider stars that are too bright for Gaia~DR3 to have a good parallax solution. We analyzed the membership probabilities of all entries within 100\,pc of the Sun using BANYAN~$\Sigma$ and rejected those with resulting Bayesian Oceanus membership probabilities below 90\%, or with best-case scenario separations of more than 3\,\kms\ between the star's $UVW$ position and the center of the Oceanus moving group model. The latter step is commonly used (e.g., see \citealp{2018ApJ...862..138G}) to exclude young stars of yet unknown moving groups that are a poor match to both the young models and the field models of BANYAN~$\Sigma$. While we have found no additional members from Hipparcos, the Gaia~DR3 search revealed an additional 74 candidate members, discussed further in Section~\ref{sec:disc}.

\subsection{Missing Substellar Members}\label{sec:submem}

We used the Ultracoolsheet \citep{zenodoultracoolsheet} that compiles the properties of most currently known brown dwarfs as well as the CatWISE2020 full-sky catalog \citep{2021ApJS..253....8M} to search for substellar candidate members in the Oceanus moving group.

We based the Ultracoolsheet search on the best available kinematic measurements listed in the table, using the photometric distances provided in the Ultracoolsheet when no parallax was available. We applied the same selection criteria as those described in Section~\ref{sec:stellar}, i.e., we required membership probabilities above 90\%\ and best-case scenario separation below 3\,\kms\ in $UVW$ space, which yielded 5 substellar candidate members listed in Table~\ref{tab:candidates}, including the Luhman~16~AB system described further in Section~\ref{sec:luh16}.

\begin{figure*}[p]
	\centering
	\subfigure[Brown dwarf (2010)]{\includegraphics[width=0.45\textwidth]{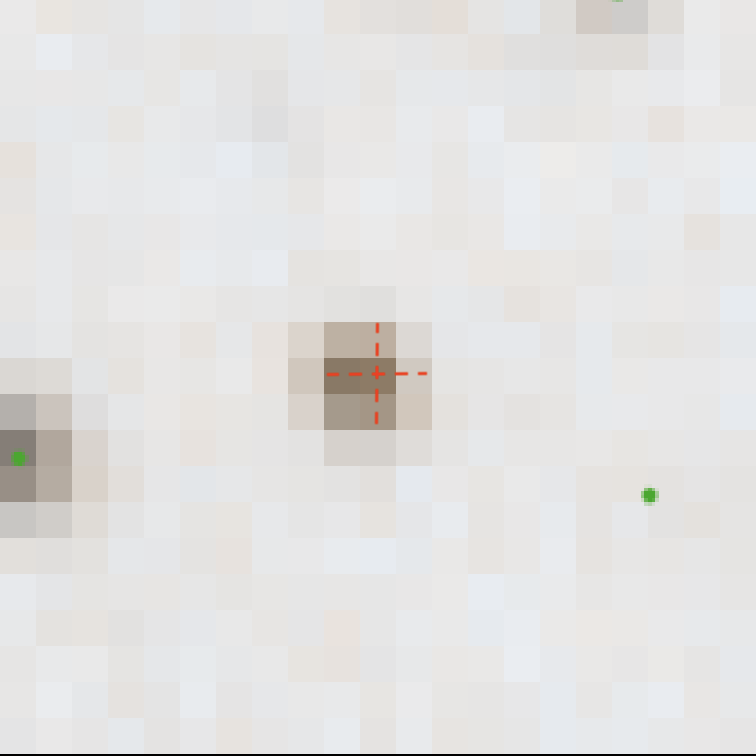}\label{fig:wv1}}
	\subfigure[Brown dwarf (2020)]{\includegraphics[width=0.45\textwidth]{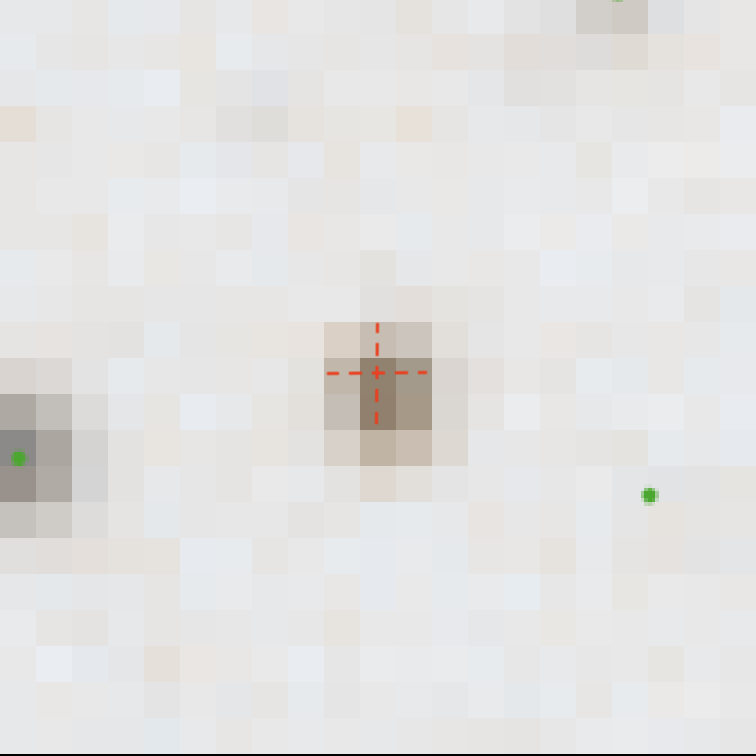}\label{fig:wv2}}
	\subfigure[False positive (2010)]{\includegraphics[width=0.45\textwidth]{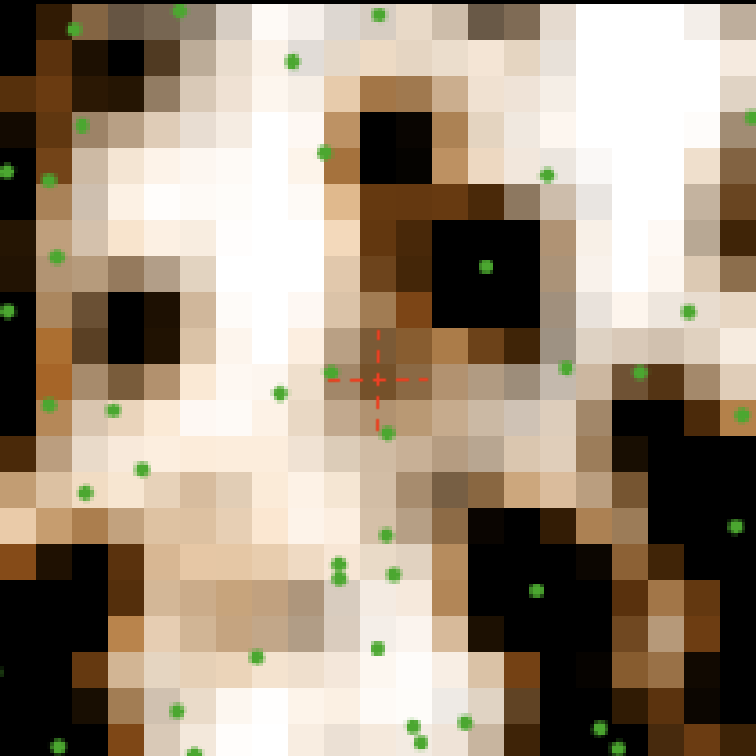}\label{fig:wv3}}
	\subfigure[False positive (2020)]{\includegraphics[width=0.45\textwidth]{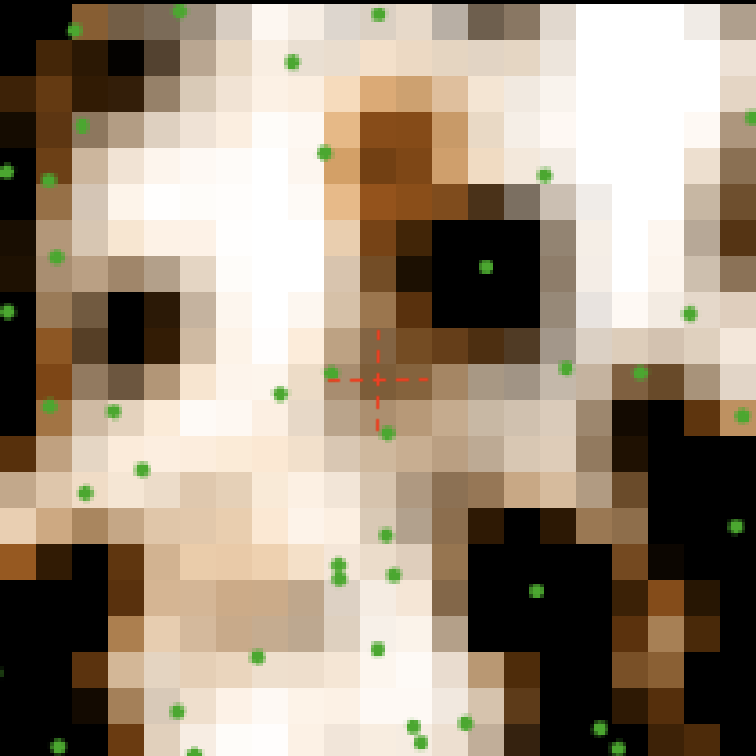}\label{fig:wv4}}
	\caption{Example WiseView 60$\times$60$^"$ $W1+W2$ snapshots of the brown dwarf CWISE~J114202.84+412337.5 and (top row) and the false-positive motion candidate CWISE~J174710.22--280815.6 (bottom row). Red crosshairs mark the CatWISE2020 catalog positions of both objects, and green circles indicate locations of nearby Gaia~DR3 entries. While the top row clearly indicates a displacement of the point spread function over the 2010--2020 decade, the bottom row shows a crowded field of view with nearby sources that show significant flux variation over time; the contamination caused by such variables induced a false detection of motion in the CatWISE2020 catalog, however, the central source is clearly not moving significantly.}
	\label{fig:wiseview}
\end{figure*}

We used the full-sky CatWISE2020 catalog to identify additional brown dwarfs that are not yet known in the Oceanus moving group, using an ADQL query on the VizieR TAP/ADQL service\footnote{Available at \url{https://tapvizier.u-strasbg.fr/adql/}.} to assemble a list of potential substellar objects with good-quality proper motion. We only considered sources with $W1-W2$ colors above 0.5\,mag consistent with a spectral types T0 or later \citep{2011ApJS..197...19K}, and with total proper motions of 100\,\masyr\ or above. We also applied a selection cut on the \texttt{w1sky} (we require a value between -5 and 5) and \texttt{w2sky} (between -5 and 10) to avoid sources where the background flux is relatively high, which we found significantly reduces the sample contamination due to background galaxies. We have determined a photometric distance for each source using the color-magnitude sequence of all known brown dwarfs in an absolute $W1$ versus $W1-W2$ color-magnitude diagram, which is useful to predict the photometric distances of brown dwarfs of spectral types T0 and later, regardless of their ages (e.g., see \citealp{2021ApJS..253....7K}). Using these photometric distances and the proper motions provided in the CatWISE2020 catalog, we determined the BANYAN~$\Sigma$ membership probabilities of all catalog objects and applied the same selection criteria as described above to identify 11 potential substellar members of the Oceanus moving group. We have visually vetted the resulting candidates in the WISE individual epoch images using the WiseView tool \citep{2018ascl.soft06004C} as shown in Figure~\ref{fig:wiseview}, and rejected 10/11 which we found to be spatially extended and photometric variable sources, and thus likely distant galaxies. The remaining single substellar candidate is CWISE~J114202.84+412337.5, with a photometric spectral type estimate of T0 based on its $W1-W2$ color \citep{2011ApJS..197...19K}. This high rate of false detections is not unexpected for such blind searches based on high motion candidates in CatWISE2020; however, these false-positive moving sources are easy to discriminate by eye or with neural networks when large numbers are involved. This is why citizen science projects such as `Backyard Worlds: Planet~9' (e.g., see \citealt{2017ApJ...841L..19K}) have proven so successful in detecting new brown dwarfs using \textit{WISE} data. Adopting a lower proper motion cut below 100\,\masyr\ would dramatically increase the false-positive rate, given that CatWISE2020 measurement errors on proper motions are relatively large (60--100\,\masyr\ for targets with $W2$ magnitudes in the range 16--19).

In addition to the searches described above, we also used the BASS survey input sample of \cite{2015ApJ...798...73G} to identify additional Oceanus candidate members, specifically to cover the full range of spectral types from mid M to mid L, which are not necessarily well covered by searches based on literature compilations or brown dwarfs or those based on CatWISE2020 proper motions alone. The BASS survey is based on a cross-match of the 2MASS \citep{2006AJ....131.1163S} and AllWISE \citep{2014ApJ...783..122K} all-sky near-infrared surveys, and identifies all objects with proper motions larger than 30\,\masyr\, colors consistent with a spectral type M5 or later, and various quality cuts. We used the method of \cite{2015AA...574A..78S} to assign photometric spectral type estimates to all entries of the BASS catalog, and we used the spectral type to absolute AllWISE $W1$ magnitudes sequence of known M dwarfs and brown dwarfs (e.g., see \citealp{2011ApJS..197...19K}) to assign a photometric distance estimate. In contrast with the previous use of absolute $W1$ magnitudes versus $W1-W2$, such photometric distance estimates would be biased for brown dwarfs significantly younger than $\approx 200$\,Myr (e.g., see \citealp{2016ApJS..225...10F}), however, the Oceanus moving group is old enough that this bias will not affect our membership probabilities. We applied the same selection criteria as described above to select good-quality candidate members, allowing us to recover 6 candidates with photometric spectral type estimates in the range M6--L8.5. A visual inspection with WiseView allowed us to reject 2/6 objects that are likely background galaxies. Two of the remaining four objects are known literature brown dwarfs, and all four objects are listed in Table~\ref{tab:candidates}. A literature search for higher-quality proper motions allowed us to further reject 3/4 as non-members of the Oceanus moving group, and only the photometric L6 candidate 2MASS~J12543853+4346573 remains as a valid candidate member.

\subsection{Membership Categories}\label{sec:memcat}

Following the convention of \cite{2004ARAA..42..685Z} and \cite{2008hsf2.book..757T}, we designate the most secure members of the Oceanus moving group as `bona fide members'. We require that bona fide members have full kinematic measurements, including a parallax and radial velocity, and at least one indication of youth. We also require that their BANYAN~$\Sigma$ probabilities be above 90\% with a separation below 3\,\kms\ from the central point of the BANYAN~$\Sigma$ model in $UVW$ space; this latter criterion is useful to reject contaminants from yet unrecognized moving groups that are not a good fit to any model currently included in BANYAN~$\Sigma$. By construction, only 80--90\% of the true Oceanus members will obtain a BANYAN~$\Sigma$ probability above 90\%\footnote{This translation between Bayesian probability and false-positive rates is only valid for all associations for Bayesian probabilities of exactly 90\%, and are association-dependent for other thresholds. See Section~7 of \cite{2018ApJ...856...23G} for a detailed explanation.}.

Candidate members that lack either an observed signature of youth or one component of their full 6-dimensional kinematics, or with a BANYAN~$\Sigma$ probability in the range 80\%--90\%, are designated as `high-likelihood candidate members'. Most of these stars will likely be true members, but still require some final observational corroboration of their membership.

Candidate members that lack more than one of the observables listed above but still have a BANYAN~$\Sigma$ probability above 80\% are designated as `candidate members'. We estimate that roughly 85\% of those with radial velocity measurements, and 66\% of those without radial velocity measurements are true members, based on past surveys of similar sparse and nearby young associations \citep{2018ApJ...862..138G}. Among the stars that show at least one sign of youth, we expect the true fraction of members to be much higher given the small fraction of young field stars in the solar neighborhood (about 20\% of nearby stars are expected to be younger than $\approx$\,1\,Gyr \citealt{2014AA...569A..13R}), which would correspond to respective true positive rates of 97\% and 93\% in our sample of stars with at least one sign of youth. However, these latter estimates are optimistic because they ignore the possible cross-contamination between young associations, especially when a radial velocity measurement is lacking. Both high-likelihood and regular candidate members are also required to be located within 3\,\kms\ from the central point of the BANYAN~$\Sigma$ model in $UVW$ space.

Candidate members with at least one problematic kinematic measurement, have a BANYAN~$\Sigma$ membership probability below 80\%, or that have a $UVW$ separation above 3\,\kms\ from the central point of the BANYAN~$\Sigma$ model, are designated as `low-likelihood candidate members'. Based on our past experience with such objects, they are composed of a mix of scenarios : (1) random interlopers from the field that initially appeared as good candidate members based on a limited set of kinematic measurements; (2) stars with unreliable kinematic measurements, sometimes due to an unresolved companion that affects the proper motion, parallax or radial velocity; or (3) contaminants from other yet unrecognized young associations.

The candidate members that present at least one clear indication that they are much older or much younger than the Oceanus moving group, or which kinematic measurements clearly preclude a match to the Oceanus moving group, are assigned the `rejected' status. With the rare exception of stars which kinematics are heavily affected by an unresolved companion, most of these objects are typically never recovered as candidate members when additional measurements are obtained.

There are occasional cases where two components of a gravitationally bound binary or a higher-order multiple get assigned to different membership categories. While this is expected to happen based on random fluctuations in measurements within the quoted measurement errors, some cases are triggered by either of the components being affected by strong systematics in their kinematic measurements. When either of the components displays a clear sign that it is too young or too old to be a member of the Oceanus moving group, we reject both components\footnote{e.g., the 2MASS~J16115252+1710171 binary system is composed of Gaia~DR3~1198870714007305472 (no radial velocity) which obtains a 98.4\% membership in the Oceanus moving group, but also of a second co-moving Gaia~DR3 entry Gaia~DR3~1198870718304793600, which has a Gaia~DR3 radial velocity measurement rejecting membership in the Oceanus moving group, and we therefore reject the whole system.}, however, when a single kinematic measurement is in cause, we have to rely on more subtle determinations of which component should be trusted more, for example based on the value of the Gaia~DR3 Renormalised Unit Weight Error (RUWE\footnote{The RUWE is documented at \url{https://gea.esac.esa.int/archive/documentation/GDR2/Gaia_archive/chap_datamodel/sec_dm_main_tables/ssec_dm_ruwe.html} and is an indication of the quality of the parallax solution.}) of both components.

This is the case with PM~J19098+0742~AB; the brighter component obtains a good membership probability in the Oceanus moving group (96.0\%), but the fainter companion obtains a much lower probability (27.7\%) and has a smaller Gaia~DR3 RUWE value and a smaller measurement error on its parallax, which means that its astrometric solution may be more reliable. We have therefore elected to keep the system in the list of regular `candidate members' until a better astrometric solution is available.

Similarly, both V2121~Cyg~A and V2121~Cyg~BC are strong radial velocity variable in Gaia DR3; V2121~Cyg~BC displays strong X-ray emission typical of ~100\,Myr-old members of the AB~Doradus moving group and obtains a high membership probability in the Oceanus moving group (99.9\%), however, the radial velocity of component A lowers its Oceanus membership probability significantly (from 99.6\% to 0.6\%). \cite{2003AJ....125.2196F} noted that the radial velocity of V2121~Cyg~A is difficult to measure because of significant line profile asymmetries in its spectrum, possibly due to the $\gamma$~Dor pulsations of this star. We have therefore also elected to keep this system in our list of regular `candidate members' until its systemic radial velocity can be assessed more robustly.

The final membership categories of all candidate members of the Oceanus moving group are listed in Table~\ref{tab:candidates}, along with their BANYAN~$\Sigma$ membership probabilities, and the list of available observables that were input to BANYAN~$\Sigma$ to determine these probabilities. In Table~\ref{tab:kin}, we list the detailed kinematics of each candidate member, mostly obtained from Gaia~DR3, with the exception of the coldest candidate members too faint for Gaia, or some of the brightest members (e.g., HD~157750 and HD~97334~B) that can still have a more precise parallax or proper motion measurement in Hipparcos \citep{2007AA...474..653V} or the Tycho-Gaia astrometric catalog \citep{2016AA...595A...2G}.

The radial velocities listed in Table~\ref{tab:kin} were combined from all the available radial velocities available in the literature (listed in Table~\ref{tab:rvs}), by a weighted average of the individual measurements, where the weights are taken as the inverse square of the measurement error, multiplied by the number of epochs the radial velocity measurement was based on. The individual measurement errors were propagated to this weighted average, and summed in quadrature with the weighted standard deviation of the individual measurements, to account for physical radial velocity variations that are not due to instrumental uncertainty. In Table~\ref{tab:xyzuvw}, we list the resulting $XYZ$ Galactic coordinates and the $UVW$ space velocities, where measurement errors were propagated with a 10$^4$-elements Monte Carlo simulation.

\section{DISCUSSION}\label{sec:disc}

\subsection{Rotation Periods}\label{sec:prots}

We used public data from the \textit{Transiting Exoplanet Survey Satellite} (TESS, \citealp{2014SPIE.9143E..20R}) to measure the rotation periods of Oceanus members and candidate members. We recovered the Full Frame Images (FFIs) on the Mikulski Archive for Space Telescopes (MAST\footnote{\url{https://archive.stsci.edu}}) for 26 stars based on their Gaia DR3 coordinates. It is suggested by the TESS team to use $G$-band magnitudes in the range 9--15\,mag to obtain a good signal-to-noise ratio, however, we attempted to analyze the light curves of all targets with $G_{mag} \leq 18$ since preliminary examinations showed that a clear signal can sometimes be retrieved from fainter objects (M.~Popinchalk et al., submitted to ApJ).

\begin{figure}
	\centering
	\includegraphics[width=0.47\textwidth]{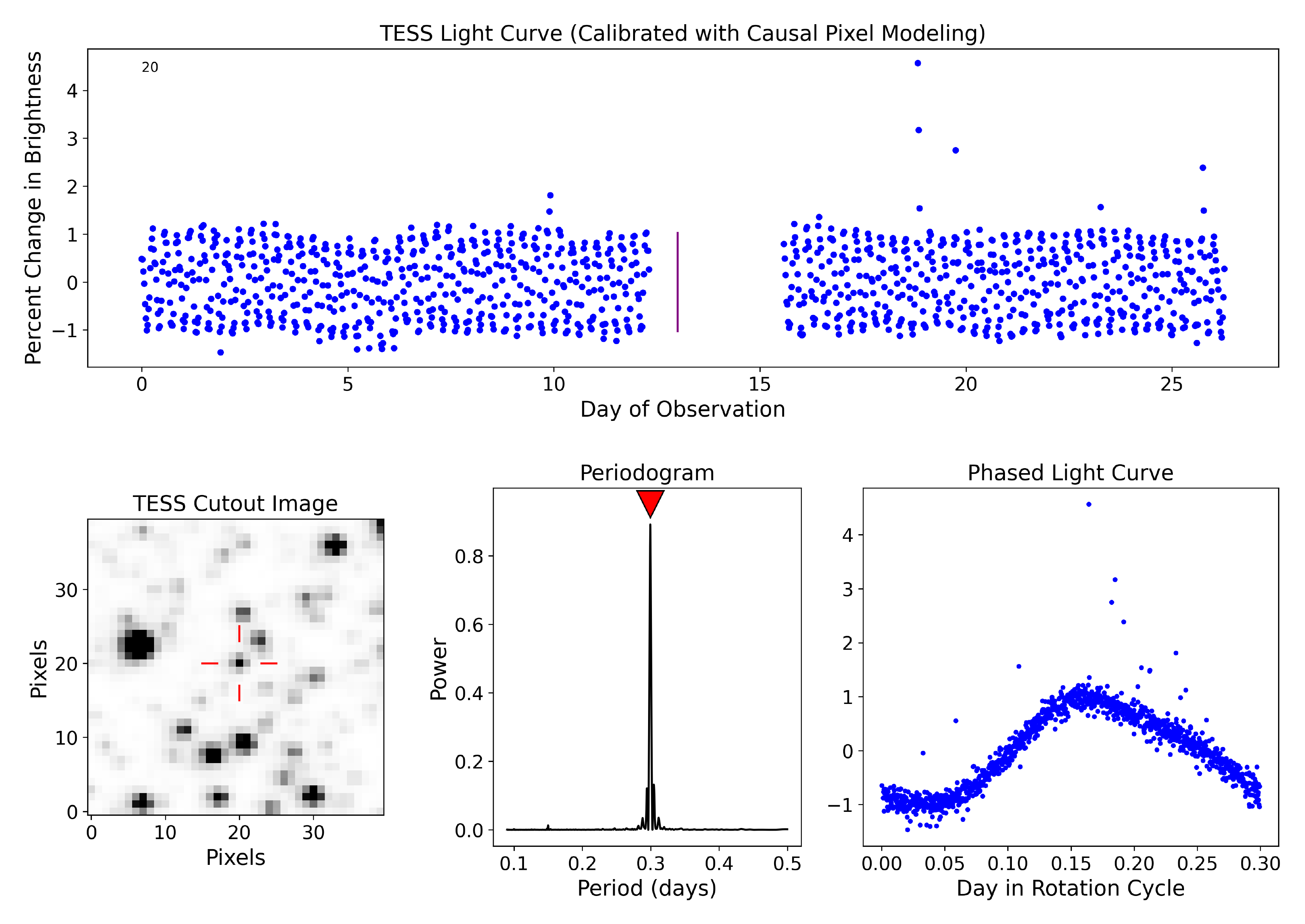}
	\caption{Example user interface figures that were used to investigate the TESS light curves. This example is for the candidate member LP~205--44, for which we measure a rotation period of 0.3\,days. The top figure displays the light curve from either the SAP or CPM pipeline. The bottom left figure displays a TESS acquisition cutout image, where each pixel covers an area of $21''$ by $21''$. The figure at the middle column of the bottom row shows a Lomb-Scargle periodigram with the best-fitting period, and the bottom right figure shows a light curve phased to the best-fitting period.}
	\label{fig:panel_example}
\end{figure}

\begin{figure}
	\centering
	\includegraphics[width=0.47\textwidth]{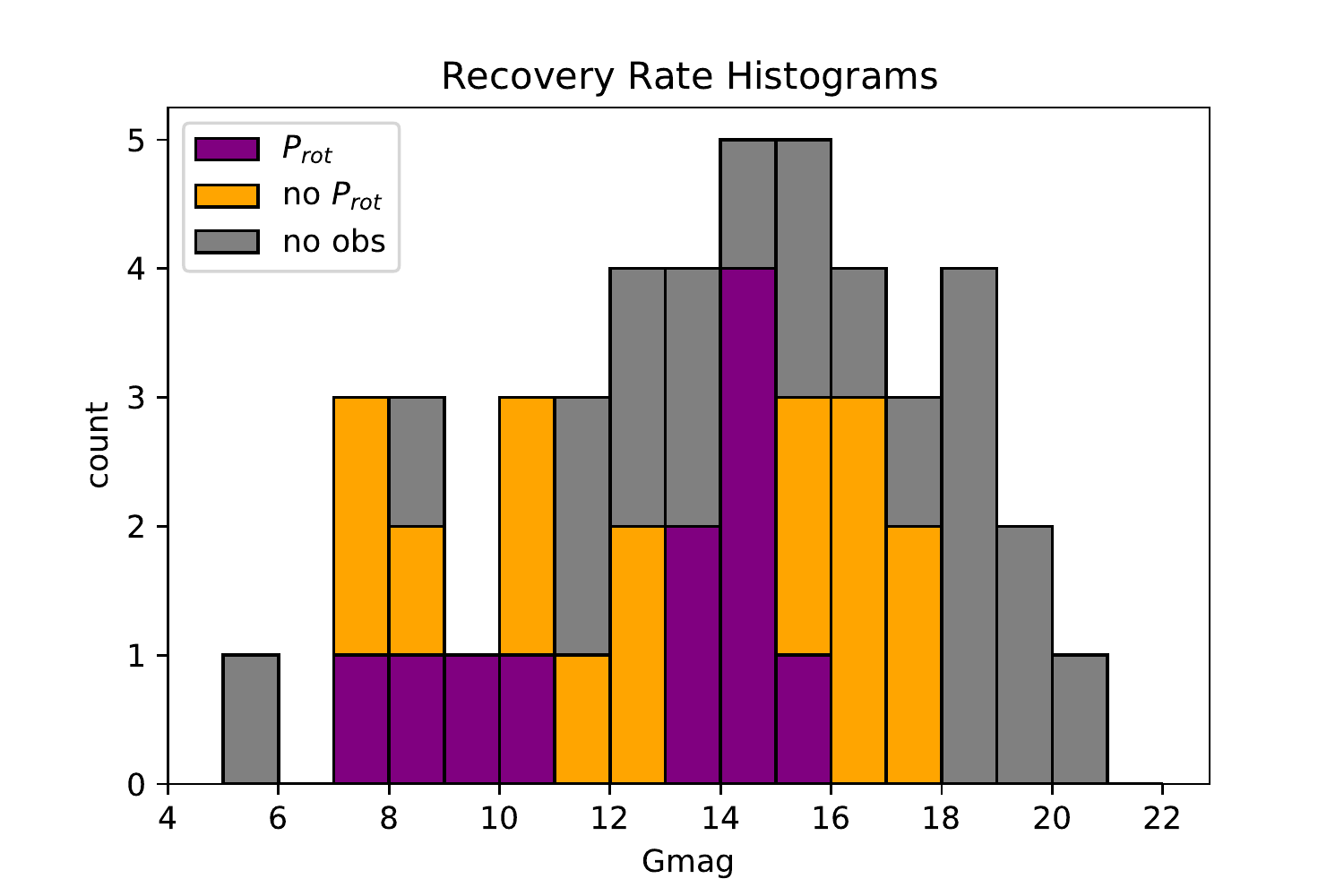}
	\caption{Distribution of members and candidate members of the Oceanus moving group for which a rotation period could be recovered with TESS (purple), no rotation period could be recovered because of problematic or low-signal to noise data (orange), or stars for which no TESS data were available (grey).} 
	\label{fig:recovery_rates}
\end{figure}

We investigated the light curve of each star and each available TESS sector, as generated from two pipelines: The Simple Aperture Photometry (SAP) for stars with $G_{mag} \leq 10$, and the Causal Pixel Model (CPM; \citealt{2022AJ....163..284H}) for stars with $G_{mag} \geq 10$. We used both pipelines for stars with $G_{mag} \approx 10$. We used the \texttt{tess\_check} Python library\footnote{Available at \url{https://github.com/SPOT-FFI/tess check}} to inspect the light curves with a user interface that includes a 4--figures panel shown in Figure~\ref{fig:panel_example}. We then visually classified the quality of each light curve, and whether it shows clear variability or a flat curve. The light curves of some stars were not considered, either because of de-trending problems, contamination from a neighbor star, or low-signal to noise data. Figure~\ref{fig:recovery_rates} shows the distribution of $G$-band magnitudes for the stars in our sample as a function of whether a rotation period could be recovered.

\begin{figure*}[p]
	\centering
	\subfigure[Prot]{\includegraphics[width=0.45\textwidth]{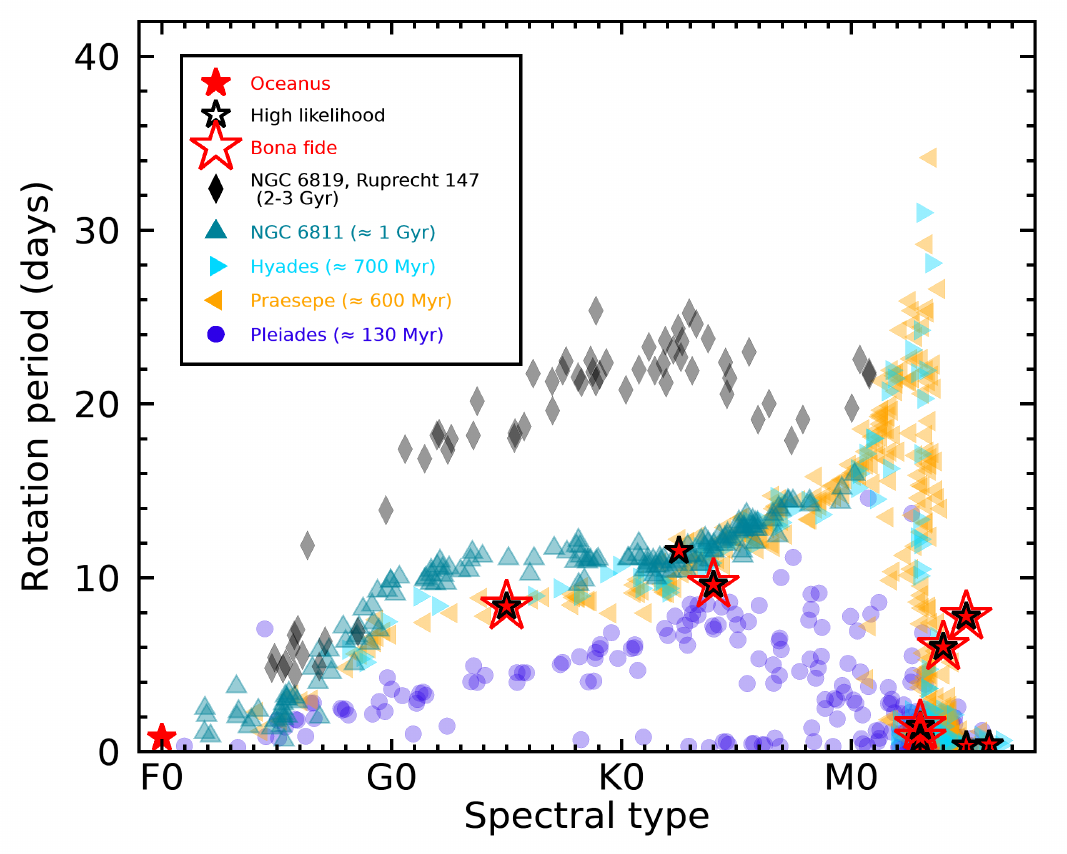}\label{fig:prots}}
	\subfigure[Activity]{\includegraphics[width=0.45\textwidth]{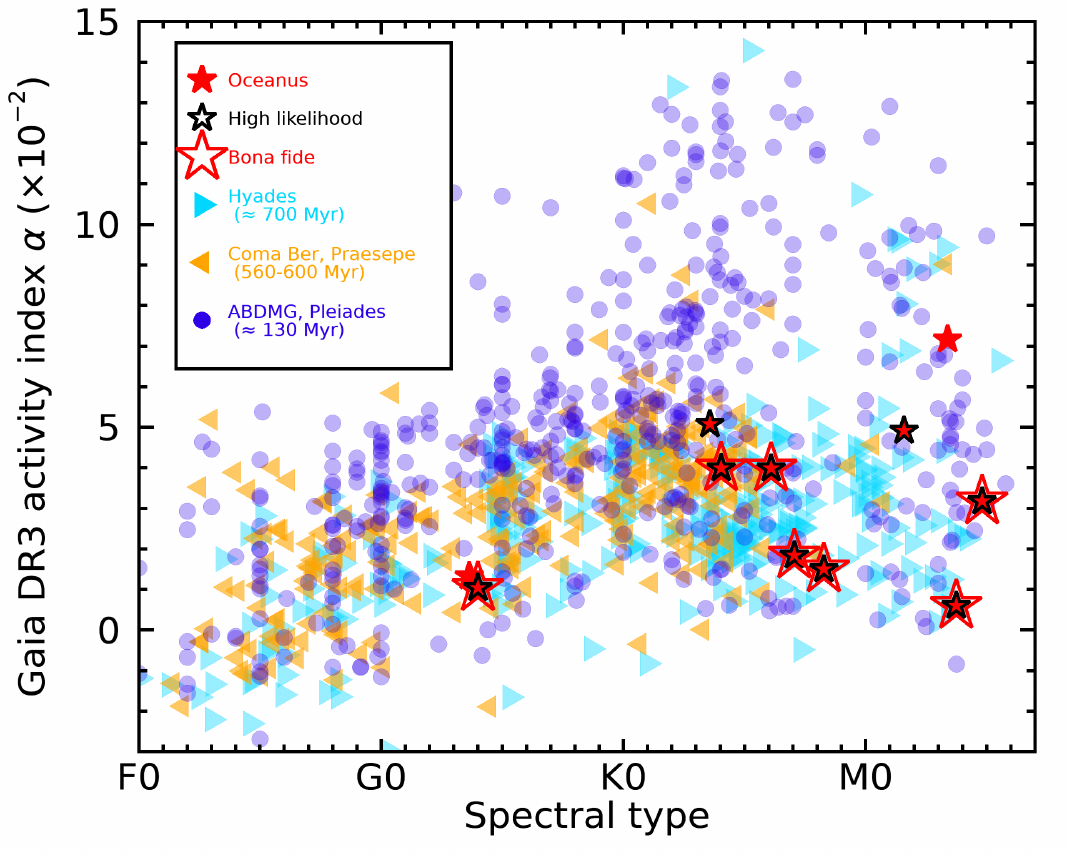}\label{fig:actind}}
	\subfigure[Galex]{\includegraphics[width=0.45\textwidth]{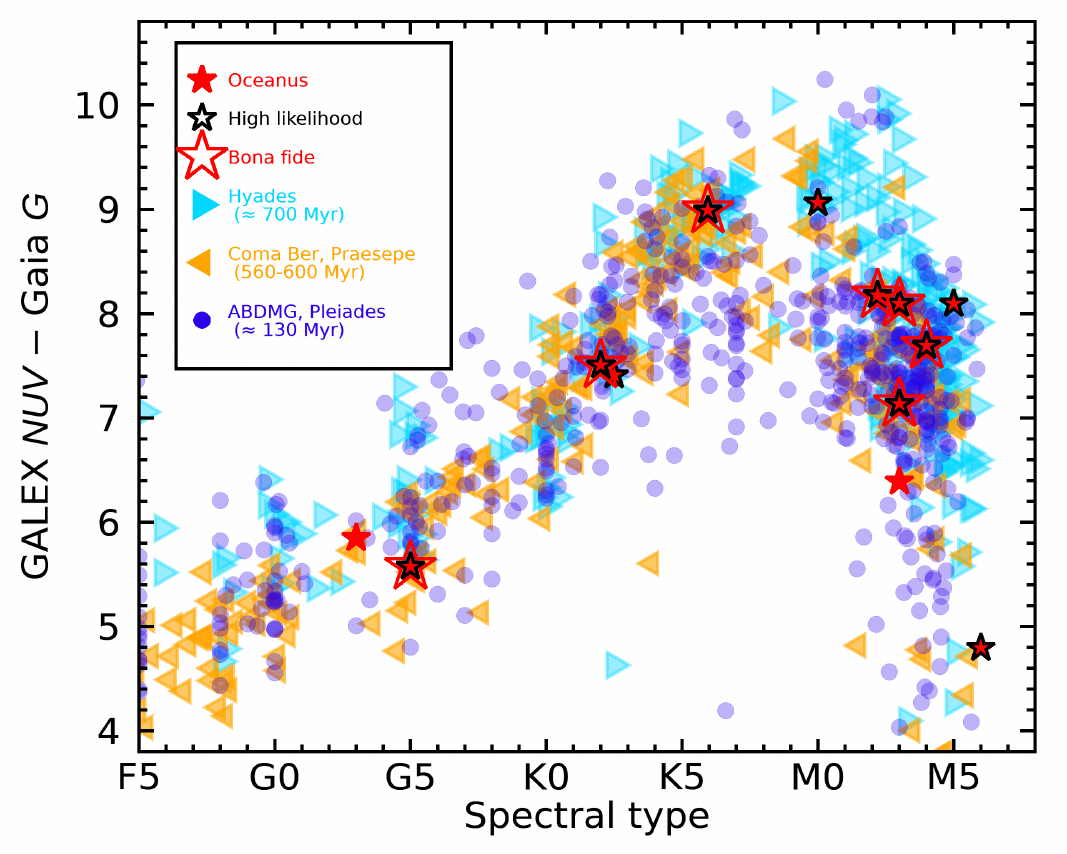}\label{fig:galex}}
	\subfigure[Xray]{\includegraphics[width=0.45\textwidth]{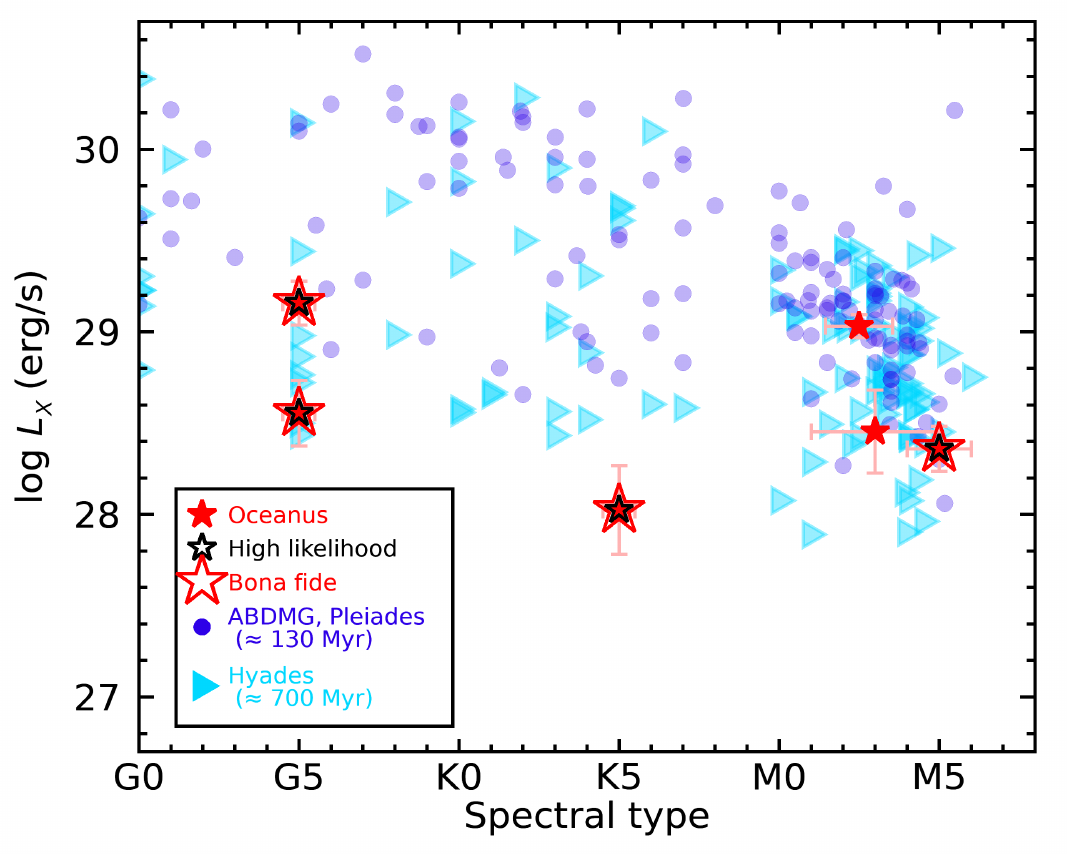}\label{fig:xray}}
	\caption{Rotation periods and activity-dependent indices for members and candidate members of the Oceanus moving group (red stars), compared with members of other young associations (colored symbols). The rotation periods are consistent with the Hyades and Coma~Ber sequences, and all other activity indices are consistent with an age in the range 560--700\,Myr. We highlighted the high-likelihood candidate members with a black, open star symbol, and the bona fide members with an additional red, open star symbol. The low-likelihood and rejected candidate members are not shown here.}
	\label{fig:act}
\end{figure*}

The resulting rotation periods are shown in Figure~\ref{fig:act}, are listed in Table~\ref{tab:youth} and are discussed further in Section~\ref{sec:age}.

\subsection{Age Determination}\label{sec:age}

We compile in Table~\ref{tab:youth} and show in Figure~\ref{fig:act} a literature compilation of stellar activity-related measurements for members of the Oceanus moving group, including $NUV - G$ colors from \textit{GALEX} \citep{2005ApJ...619L...1M} and Gaia~DR3, X-ray luminosity from the ROSAT mission \citep{1982AdSpR...2d.241T}\footnote{We queried the ROSAT All Sky Survey Faint Source Catalog \citep{2000IAUC.7432....3V} and the ROSAT All-Sky Bright Source Catalogue \citep{1999AA...349..389V}}, and the newly available Gaia~DR3 chromospheric activity index \texttt{astrophysical\_parameters.activityindex\_espcs} (also called $\alpha$), which is a spectral index measured on the calcium triplet of the Gaia RVS spectra that scales with stellar activity \citep{2022arXiv220605766L}\footnote{See also \url{https://gea.esac.esa.int/archive/documentation/GDR3/Data_analysis/chap_cu8par/sec_cu8par_apsis/ssec_cu8par_apsis_espcs.html} for more detail.}. Taken together with the distribution of rotation periods shown in Figure~\ref{fig:prots}, members of the Oceanus moving group are globally consistent with a coeval population of an age significantly older than the Pleiades ($127.4_{-10}^{+6.3}$; \citealp{2022AA...664A..70G}), but younger than NGC~6811 ($\approx$\,1\,Gyr; \citealp{2011ApJ...733L...9M}). Furthermore, they are globally consistent with the members of Coma Ber ($562^{+98}_{-84}$\,Myr; \citealp{2014AA...566A.132S}), Praesepe ($617.0_{-10}^{+40}$; \citealp{2018ApJ...863...67G}) and the Hyades ($695_{-67}^{ +85}$; \citealp{2022AA...664A..70G}), but do not clearly distinguish between either case. It is therefore immediately apparent from these activity indices that the age of the Oceanus moving group is likely much older than 130\,Myr, and possibly as old as about 700\,Myr.

Using the gyrochronology relations of \cite{2008ApJ...687.1264M} for the three members in the appropriate range of spectral types for which we have rotation periods, we obtain ages of $360_{-110}^{+150}$\,Myr for the K dwarf TYC~3424--1000--1 (its Gaia $G-GRP$ color is consistent with a spectral type K4); $520_{-75}^{+90}$\,Myr for the G5 star HD~207485; and $620_{-80}^{+90}$\,Myr for the K2.5 star YY~LMi. Because TYC~3424--1000--1 lacks a spectral type or high-quality photometric measurements in the $B$ band (literature measurements span 12--12.4\,mag; \citealp{2004AAS...205.4815Z,2015AJ....150..101Z}), we choose to use only the gyrochronology ages of YY~LMi and HD~207485 to estimate the age of the Oceanus moving group at $560 \pm 60$\,Myr by combining their age probability density functions. The age estimate of TYC~3424--1000--1 is marginally consistent with the age of the Oceanus moving group, but a spectroscopic follow-up of this star will be required to refine its gyrochronology.

\begin{figure*}
	\centering
	\includegraphics[width=0.95\textwidth]{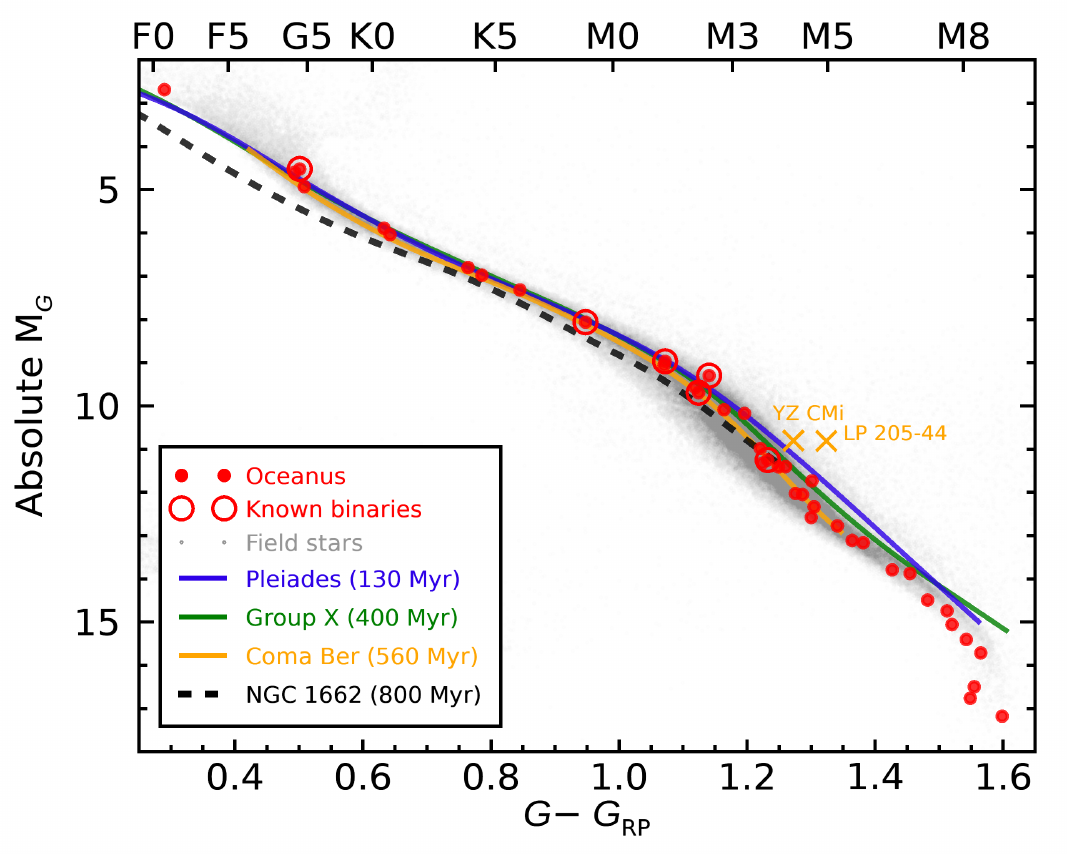}
	\caption{Gaia~DR3 color-magnitude diagram of Oceanus members and candidate members (filled red circles), with known binaries highlighted with larger, open red circles. Empirical sequences are shown for four other associations with known ages. Members of the Oceanus moving group fall along the Coma~Ber sequence, indicating that they are likely coeval with this open cluster.}
	\label{fig:cmd}
\end{figure*}

A color-magnitude diagram of the Oceanus moving group members is shown in Figure~\ref{fig:cmd}, and compared with best-fit sequences to members of different young associations (Pleiades, $127.4^{+6.3}_{-10}$\,Myr, \citealt{2022AA...664A..70G}; Group~X, $400 \pm 80$\,Myr, \citealt{2017AJ....153..257O} and \citealt{2022AA...657L...3M}; Coma~Ber, $562^{+98}_{-91}$\,Myr, \citealt{2014AA...566A.132S}; and NGC~1662, $800 \pm 160$\,Myr, \citealt{2020AA...640A...1C}) constructed with the method of \cite{2021ApJ...915L..29G}. This figure clearly demonstrates that the early-type members of the Oceanus moving group are inconsistent with an age of 800\,Myr or older, but its late-type members appear older than the $400 \pm 80$\,Myr-old members of \cite{2017AJ....153..257O} Group~X. The empirical best-fit sequence to members of Coma~Ber provide a very good match to the color-magnitude sequence of the Oceanus moving group members, consistent with our age determinations based on the \cite{2008ApJ...687.1264M} gyrochronology relations.

We performed a quantitative isochronal age determination of the Oceanus moving group by comparing the empirical sequences of Group~X, Coma~Ber and NGC~1662 (as shown in Figure~\ref{fig:cmd}), restrained to members of Oceanus that were not flagged as likely binaries, and to the range of Gaia~DR3 colors where all empirical sequences are defined (i.e., $0.493 < G-G_{\rm RP} < 1.249$), leaving us with 17 high-likelihood or bona fide members and 2 additional regular candidate members for the isochronal age determination. We used a Markov Chain Monte Carlo algorithm with the DREAM-ZS sampler including a snooker updater (see \citealt{terBraak:2008iw} for more details, and \citealt{2016ApJ...819..133A} for a use case example) with 12 chains and $10^4$ Monte Carlo iterations (the first third ignored as a burn-in phase), with a log likelihood expression that assumes Gaussian error bars on the absolute $G$--band magnitudes and Gaia~DR3 $G-G_{\rm RP}$ colors, projected to the $Y$ axis using the local derivative of the model sequence being tested. The Markov Chain Monte Carlo has a single age parameter, which is used to interpolate between empirical sequences in log age. We use a simple Gaussian likelihood similar to a classical $\chi^2$, with a flat prior on log age. We obtained our final isochronal age estimates by taking the values at 50\%\ of the cumulative distribution function of all samples for the central value, and at 15.9\%\ and 84.1\%\ for the 1$\sigma$ measurement errors.

We obtained an age of $520 \pm 30$\,Myr or $510 \pm 30$\,Myr, respectively, by using all members, or only high-likelihood and bona fide members of the Oceanus moving group, also consistent with other age determinations described above. The $30$\,Myr measurement error only reflects the measurement errors of the photometry of Oceanus members and candidate members, and the respective standard deviations of open cluster members around their best-fit sequences. We add an additional systematic uncertainty of $90$\,Myr on this estimate, representative of the age uncertainties of the reference open clusters used here. Therefore, we adopt an isochronal age of $510 \pm 95$\,Myr based on only high-likelihood and bona fide members, as the most robust age indicator for the Oceanus moving group yet, and we therefore adopt it until more data is available.

This age estimate is also consistent the age of 600--800\,Myr estimated by \cite{2017ApJ...846...97G} based on a comparison of the temperature and dynamical masses of Luhman~16~AB with the \cite{2008ApJ...689.1327S} substellar evolutionary models. Our age sits on the younger end of this range, which may provide a useful constraint for evolutionary substellar models, especially once we obtain further age diagnostics for the Oceanus moving group.

Based on our age estimation of the Oceanus moving group, we could expect the lithium depletion boundary for this group to be located at spectral types L2--L3, according to the \cite{2015AA...577A..42B} substellar evolutionary models, which best match the substellar Li depletion boundary of the Hyades to other age determinations of the cluster (e.g., see \citealp{2018ApJ...856...40M}). It is therefore plausible that measuring the Li absorption of the four candidate members near this range of spectral types\footnote{Specifically, the L2 dwarf 2MASS~J16573454+1054233, and the likely L0--L1 dwarfs Gaia~DR3~1827398564574481152, Gaia~DR3~4048655804979383680 and Gaia~DR3~4131855643673923456, based on the photometric brown dwarf classification method of \cite{2015AA...574A..78S}.} may provide an additional constraint on the group's age.

It is interesting to note that we may expect a \shortage\,Myr-old population of stars to still contain A-type stars on the main sequence, whereas the earliest-type candidate member we recovered in the Oceanus moving group is the F0--type V2121~Cyg. We also recovered only three G-type candidate members (HD~207485, HD~157750, and HD~77006~A). This number of AFG-type members is comparable to similarly-sized sparse young associations such as Volans-Carina, which has 8/65 members in this range of spectral types \citep{2018ApJ...865..136G}, when accounting for the fact that the Oceanus moving group has only 43 members and candidate members that would have been recovered in the B8--M9 sample of \cite{2018ApJ...865..136G} at the distance of Volans-Carina. Therefore, the lack of A-type members in the Oceanus moving group may be a simple consequence of the small number statistics due to the limited size of this group.

\subsection{Discussion of Individual Objects}\label{sec:stars}

In this section, we discuss individual members or candidate members of the Oceanus moving group that deserve further attention.

\subsubsection{Luhman~16~AB}\label{sec:luh16}

The only missing kinematic measurement for Luhman~16~AB in the Ultracoolsheet is its heliocentric radial velocity, although a few measurements have been attempted in the literature. We have compiled these measurements in Table~\ref{tab:luhmanrvs}, to which we added two more measurements based on FIRE spectra obtained with the same method as described in \cite{2014ApJ...790...90F}, with the exception that the two AB components were not resolved and were both placed in the spectrograph slit. We used the method of \cite{2017ApJS..228...18G} to obtain heliocentric radial velocity measurements by fitting the BT-Settl substellar atmosphere models \citep{2012RSPTA.370.2765A} to the observed spectrum in the range 1.51--1.5535\,$\mu$m with the amoeba Nelder–Mead downhill simplex algorithm, which was found to produce good-quality radial velocities across a wide range of spectral types in the L and T classes (see also \citealp{2015ApJ...808L..20G,2017ApJ...841L...1G,2018ApJ...854L..27G}).

The available center-of-mass radial velocity measurements span a relatively wide range of values, between 14.0 and 22.4\,\kms, with an average of $17.6 \pm 1.6$\,\kms. This average radial velocity measurement is consistent with the radial velocity predicted by BANYAN~$\Sigma$, assuming membership in the Oceanus moving group ($16.1 \pm 0.6$\,\kms), and yields a membership probability of 91.6\%. The most likely explanation for the wide range of measured radial velocities is the presence of systematic effects, especially given that companions around Luhman~16~AB have been ruled out down to Neptune masses at periods larger than one year by a Hubble astrometric follow-up \citep{2017MNRAS.470.1140B}. Absolute heliocentric radial velocity measurements of brown dwarfs rely on atmosphere models which include complex chemistry and incomplete line lists, and these particular measurements rely on a heterogeneous set of methods, which may rely on different wavelength ranges with different fractions of line completeness in current substellar models. A long-term radial velocity follow-up of Luhman~16~AB would be useful to elucidate the reason for this wide range of literature radial velocities.

\subsubsection{51~Ari}\label{sec:ari}

The bright and nearby ($21.19 \pm 0.01$\,pc) G8 star 51~Ari was recovered in our HDBSCAN analysis as one of the founding members of the Oceanus moving group. However, the automated outlier rejection included in our construction of the BANYAN~$\Sigma$ model described in Section~\ref{sec:addmem} excluded it because it is located at the edge of the $UVW$ distribution of the Oceanus moving group. 51~Ari was observed with high-resolution visible spectroscopy by \cite{2005PASJ...57...45T}, and no lithium was detected in its spectrum (with an upper limit of 3\,m\AA), consistent with an age of at least 200\,Myr. A rotation period of 28\,days and a \logrhk\ measurement of -4.905\,dex \citep{2010ApJ...725..875I} yield respective age estimates of $4.4_{-0.7}^{+0.8}$\,Gyr and $\approx 4.8$\,Gyr \citep{2008ApJ...687.1264M}, which is inconsistent with other members of this moving group, suggesting that 51~Ari is likely a kinematic outliers unrelated to this group.

\subsubsection{YZ~CMi}\label{sec:yzcmi}

At $5.989 \pm 0.001$\,pc, YZ~CMi (GJ~285) is the nearest star that was recovered as part of the initial HDBSCAN list of Oceanus members, and remains the second nearest of all candidate members after Luhman~16~AB. YZ~CMi is a well-studied M5 fast-rotating (2.776\,days; \citealp{2019AA...621A.126D}) low-mass star with X-ray emission \citep{1999AA...349..389V} and unusually strong H-alpha emission for its spectral type \citep{2021AA...652A..28L} that is consistent with an age younger than approximately 1\,Gyr. YZ~CMi was the subject of an extended simultaneous photometric and chromatic radial-velocity followup by \cite{2020AA...641A..69B}.

YZ~CMi is a known stellar binary candidate based on a study of proper motion anomalies between Hipparcos and Gaia \citep{2019AA...623A..72K}, and has a Gaia~DR3 RUWE value of 1.35, close to the 1.40 threshold of \cite{2021ApJ...907L..33S} that is a good indicator for binary stars. The Gaia~DR3 catalog detected a small but statistically significant radial velocity variation with an amplitude of 1.9\,\kms\ from 20 epochs that span 2.1\,years, and the short-term radial velocity follow-up of \cite{2020AA...641A..69B} only detected radial velocity variations of approximately 0.15\,\kms\ in amplitude with 27 epochs that span 0.6\,years. The binary hypothesis is hard to reconcile with such stable radial velocities, given that \cite{2007ApJ...670.1367L} have also not detected a visual companion beyond the saturation radius of 0\farcs5. Even a brown dwarf companion at a projected separation of 0\farcs5 or less would be expected to have an orbital period shorter than $\approx 12$\,years, and by extension the radial velocity amplitude observed by Gaia~DR3 would translate to a radial velocity shift of $\approx 730$\,\ms\ over 0.6\,years, which was clearly not observed by \cite{2020AA...641A..69B}.

YZ~CMi is unexpectedly bright compared with other members and candidate members of the Oceanus moving group in a Gaia~DR3 color-magnitude diagram (it is brighter by 1.05\,mag, see Figure~\ref{fig:cmd}), meaning that it would need to be an unresolved triple star to explain this difference. This is a tempting explanation given its slightly high RUWE and its proper motion anomaly. However, the lack of a strong radial velocity signal uncorrelated with the chromatic index of \cite{2020AA...641A..69B}, and the lack of a detectable close companion in \cite{2007ApJ...670.1367L}, make this scenario unlikely. The Gaia~DR3 $G-G_{\rm RP}$ color of YZ~CMi appears normal for an M5 dwarf, excluding the possibility that YZ~CMi is redder rather than over-luminous in the Gaia~DR3 color-magnitude diagram. The most likely explanation is therefore that YZ~CMi is much younger than the Oceanus moving group, and for this reason we elected to downgrade the membership of YZ~CMi from `bona fide' to `low-likelihood candidate member' until this is resolved.

YZ~Cmi was claimed to be a member of the `Pleiades moving group' of \cite{1975PASP...87...37E} by \cite{2001MNRAS.328...45M}, however, this group was subsequently found to be composed of stars with such a wide range of space velocities, ages and compositions, that it is likely not physical (e.g., see \citealp{2016IAUS..314...21M}). The discovery of new loose, nearby associations such as the Oceanus moving group, with only a few members bright enough to be studied kinematically in the pre-Gaia era may provide a natural explanation for the spurious discovery of moving groups based on a small number of young stars with discrepant kinematics.

YZ~CMi was also claimed to be a candidate member of the $\beta$~Pic moving group ($\beta$PMG; see \citealp{2001ApJ...562L..87Z}) by \cite{2015AA...583A..85A}, however, recent Gaia~DR3 kinematics are inconsistent with this hypothesis and place it at more than 10\,\kms\ from the locus of $\beta$PMG, making it very unlikely to be a member of this moving group despite its location in a Gaia~DR3 color-magnitude diagram matching the age of $\beta$PMG.

\subsubsection{LP~205--44}

Much like YZ~CMi, the M5 dwarf LP~205--44 is significantly over-luminous in the Gaia~DR3 color-magnitude diagram shown in Figure~\ref{fig:cmd}. However, unlike YZ~CMi, LP~205--44 is an unambiguous radial velocity variable in Gaia~DR3 data, with a robust amplitude of 59.7\,\kms\ (with an astrometric excess noise of 0.17\,mas), but it appears over-luminous by a significant 1.86\,mag in the Gaia color-magnitude diagram. Its Gaia~DR3 $G - G_{\rm RP}$ color (1.32\,mag) appears normal for its spectral type, which means its absolute $G$-band magnitude is more consistent with members of the $\approx 50$\,Myr-old Tucana-Horologium association \citep{2000ApJ...535..959Z,2022AA...664A..70G} even after dividing its absolute flux by a factor two, despite its apparent kinematic match to the Oceanus moving group. For this reason, we demoted its membership probability to `low-likelihood candidate member' until this is resolved.

LP~205--44 was claimed as a candidate member of the AB~Doradus moving group by \cite{2012AJ....143...80S}, and a candidate member of the Hyades tidal tail by \cite{2019AA...621L...2R}. However, its Gaia~DR3 kinematics are inconsistent with either of these possibilities, unless the Gaia~DR3 median radial velocity, based on 17 epochs that span 2.7 years, is significantly different from its systemic heliocentric radial velocity. If it were a member of the Hyades, there would be similar tension with its Gaia~DR3 color-magnitude position and its age. If we adopt the radial velocity $4.4 \pm 0.1$\,\kms\ from \citep{2020AA...636A..36L} based on 55 epochs, we find that the kinematics of LP~205--44 do not match any currently known young association in BANYAN~$\Sigma$.

\subsubsection{White Dwarfs}\label{sec:wds}

Our search for additional members has recovered four known white dwarfs as potential candidate members of the Oceanus moving group. Three of them (PM~J15342+0218; 2MASS~J16042049--1331235 and UCAC4~616--058177) are clearly too old to belong to the Oceanus moving group given their Gaia~DR3 color-magnitude diagrams consistent with a total age 2.5\,Gyr or older. The DA5.5 white dwarf PM~J15342+0218 is also one of the rare white dwarfs that benefits from a literature measurement of its heliocentric radial velocity that accounts for the gravitational redshift correction ($37.4 \pm 3.8$\,\kms; \citealp{2017MNRAS.469.2102A}), a factor particularly important for white dwarfs (50.7\,\kms\ in this particular case). Including this corrected radial velocity measurement in BANYAN~$\Sigma$ also precludes a membership of PM~J15342+0218 in the Oceanus moving group.

The other white dwarf which we recovered as a candidate member of the Oceanus moving group is a hot DA1.5-type white dwarf (RX~J1727.6--3559) which can be age-dated using the \texttt{wdwarfdate} Python package \citep{2022AJ....164...62K} based on an estimation of its main-sequence lifetime and its cooling time that rely on the \cite{2018ApJ...866...21C} initial to final mass relations, the MESA stellar evolution models \citep{2016ApJ...823..102C} and the white dwarf cooling tracks of \cite{2020ApJ...901...93B}. We used the effective temperature ($31670\pm960$\,K) and surface gravity ($9.08\pm0.03$) from \citet{2021MNRAS.508.3877G} to estimate the total age of the object. However, we find a total age of $220 \pm 20$\,Myr for this white dwarf, which is significantly younger than our age estimates for the Oceanus moving group. \texttt{wdwarfdate} works under the assumption of single star evolution, which is valid approximately for a white dwarf mass in the range $0.45-1.1$\,\msun. However, the estimated mass of this white dwarf is $1.24\pm0.01$\,\msun, which could be the result of a merger and could affect the estimated age. While it is possible that this is instead a kinematic interloper, such young white dwarfs are less common and it would therefore be interesting to study this white dwarf in more details to elucidate this discrepancy, and determine whether it should be considered a member of the Oceanus moving group. 

\subsubsection{Other Brown Dwarfs}\label{sec:bds}

Apart from Luhman~16~AB, we identified 7 additional substellar candidate members (L2 or later) in the Oceanus moving group. None of them are later than T8, which we estimate as the boundary between brown dwarfs and isolated planetary mass objects at the age of this group, however, this does not preclude the existence of such late-type members because the census of Y dwarfs is incomplete even within 8\,pc of the Sun \citep{2012ApJ...753..156K}. ULAS~J151637.89+011050.1 (T6; \citealp{2013MNRAS.433..457B}) and WISE~J174640.78--033818.0 (T6.5; \citealp{2013ApJS..205....6M}) are the latest-type candidate members of the Oceanus moving group but still require radial velocity measurements to corroborate their memberships. Following up the brown dwarf members of the Oceanus moving group with time-resolved photometry could be useful given that younger brown dwarfs tend to display larger-amplitude variations (e.g., see \citealp{2020AJ....160...38V}), and combined with projected rotational velocity measurements, could lead to constraints on their radii that could in turn obtain upper limits on their ages.

\subsection{Comparison with Nearby Associations}\label{sec:asso}

\begin{figure*}
	\centering
	\subfigure{\includegraphics[width=0.45\textwidth]{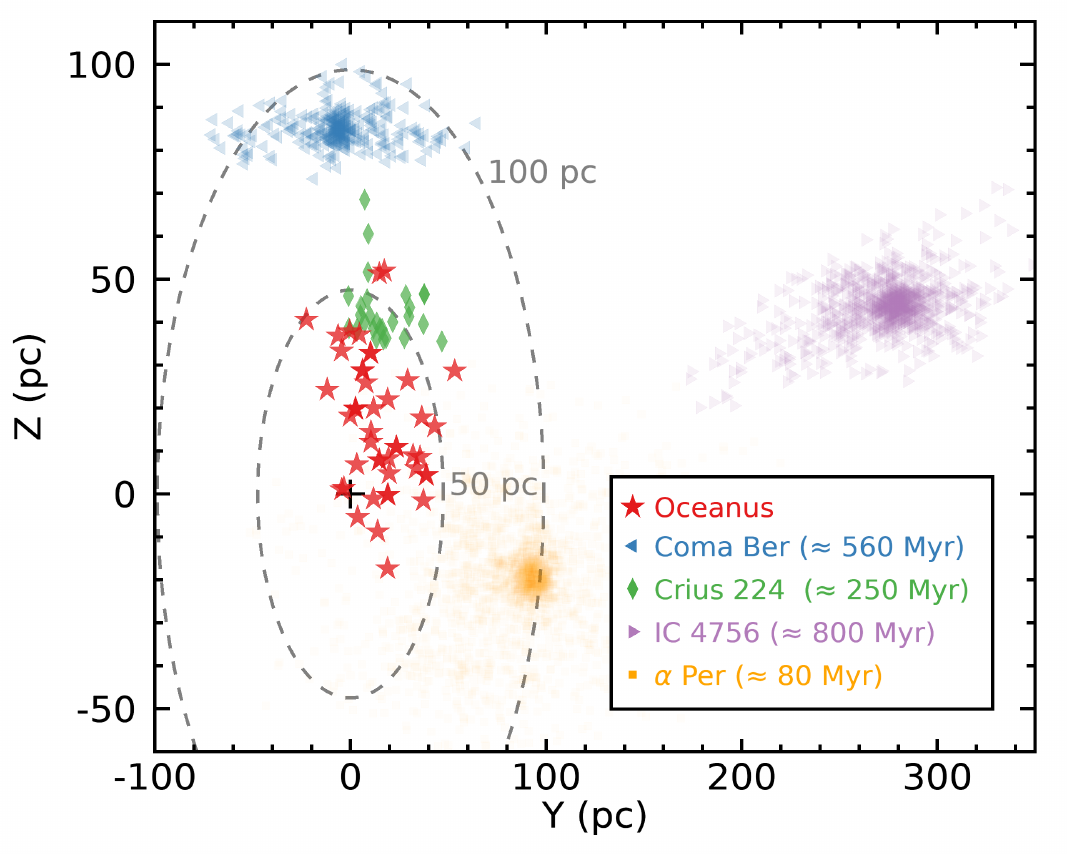}\label{fig:yz}}
	\subfigure{\includegraphics[width=0.45\textwidth]{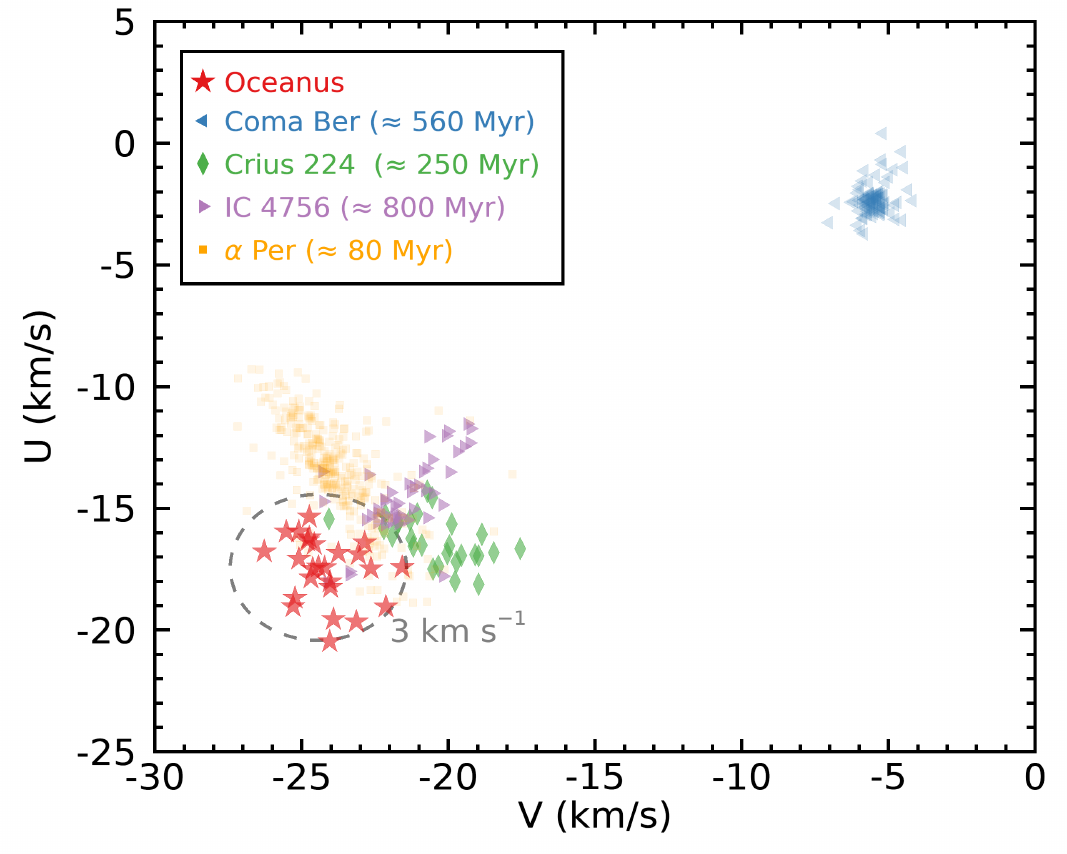}\label{fig:uv}}
	\caption{$YZ$ Galactic coordinates (left panel) and $UV$ space velocities (right panel) for members and candidate members of the Oceanus moving group (red stars) compared with other nearby young associations of similar ages or space velocities. Although IC~4756 appears potentially related to the Oceanus moving group, its older age and far distance in both the $X$ and $Y$ directions are inconsistent with its known tidal tails spatial distribution \citep{2021AJ....162..171Y}.}
	\label{fig:xzuv}
\end{figure*}

We show in Figure~\ref{fig:xzuv} projections of the spatial and kinematic distribution of the members of the Oceanus moving group, along with other nearby young associations that either have a similar age (Coma~Ber), or comparable space velocities (Crius~224 of \citealt{2022arXiv220604567M}, IC~4756, and the $\alpha$~Persei cluster). It is unlikely that Crius~224 or $\alpha$~Per are related to the Oceanus moving group given their much younger ages, and Coma~Ber is located at a much higher $Z$ Galactic coordinate, also making it unlikely that it bears any relationship to the Oceanus moving group. The only contender for both similar kinematics and a somewhat comparable age would be the $\approx$\,800\,Myr-old open cluster IC~4756 \citep{2018AA...615A.165B,2021AJ....162..171Y}. However, the age difference and the orientation of IC~4756 in the $XY$ Galactic plane and its distance make it unlikely that the Oceanus moving group is related to its tidal tails \citep{2021AJ....162..171Y}. It therefore appears more likely that members of the Oceanus moving group formed in isolation (similarly to the Volans-Carina association), although it could be related to a yet unidentified set of other moving groups that may have formed in the same environment, like the case of the TW~Hya association of \cite{1997Sci...277...67K} and the larger Sco-Cen complex.

One star that was clustered along with Crius~224 by \cite{2022arXiv220604567M}, Gaia~DR3~1484502295643717888, appears to have kinematics that are a better match to the Oceanus moving group, but an analysis based on BANYAN~$\Sigma$ makes it a poor match to either association, and rather a candidate member of the more loosely defined Cas-Tau association of young stars \cite{1956ApJ...123..408B} that is related to the $\alpha$~Persei cluster.

\subsection{Exoplanets}\label{sec:exo}

A cross-match between the Gaia~DR2 identifiers of all members and candidate members of the Oceanus moving group with the confirmed exoplanets and the TESS Project Candidates at the NASA exoplanet archive\footnote{Available at \url{https://exoplanetarchive.ipac.caltech.edu}.} returned no results. It is therefore likely that there are currently no known exoplanets or exoplanet candidates around any member or candidate member of the Oceanus moving group.

\section{CONCLUSIONS}\label{sec:conclusion}

We present the discovery of a new relatively young and loose moving group very close to the Sun that was previously not identified due to its modestly young age and low spatial density. This discovery was allowed by the pristine kinematics of Gaia~DR3 that allowed us to identify over-densities directly in 6-dimensional Galactic positions and space velocities, and make it possible to recognize even loose young associations despite geometric projection effects at distances within 100\,pc of the Sun.

We compiled a list of likely members in this new association and identified 8 likely brown dwarf members using data from CatWISE2020 and literature compilations of brown dwarfs, including the very well-studied brown dwarf binary Luhman~16~AB at 2\,pc from the Sun that was previously suspected to be relatively young.

Using new rotation period measurements from TESS and a set of literature activity indices and a Gaia~DR3 color-magnitude diagram, we show evidence that this new association is coeval and determine a gyrochronology age of \age\,Myr. This association will be a valuable laboratory to study the properties of modestly young brown dwarfs down to the coolest temperatures due to its proximity, and its low-mass members will also be targets for exoplanet searches.

\acknowledgments

We thank our anonymous reviewer for their detailed recommendations that significantly improved the quality of this paper. We would also like to thank Tim Bedding for useful comments. This project was developed in part at the Gaia Fête, hosted by the Flatiron Institute Center for Computational Astrophysics in 2022 June. This work has benefited from The UltracoolSheet at \url{http://bit.ly/UltracoolSheet}, maintained by Will Best, Trent Dupuy, Michael Liu, Rob Siverd, and Zhoujian Zhang, and developed from compilations by \cite{2012ApJS..201...19D}, \cite{2013Sci...341.1492D}, \cite{2016ApJ...833...96L}, \cite{2018ApJS..234....1B}, and \cite{2021AJ....161...42B}. 

This research has made use of the NASA Exoplanet Archive, which is operated by the California Institute of Technology, under contract with the National Aeronautics and Space Administration under the Exoplanet Exploration Program. This paper includes data collected with the TESS mission, obtained from the MAST data archive at the Space Telescope Science Institute (STScI). Funding for the TESS mission is provided by the NASA Explorer Program. Resources supporting this work were provided by the NASA High-End Computing (HEC) Program through the NASA Advanced Supercomputing (NAS) Division at Ames Research Center for the production of the SPOC data products. STScI is operated by the Association of Universities for Research in Astronomy, Inc., under NASA contract NAS 5–26555. This research made use of the SIMBAD database and VizieR catalog access tool, operated at the Centre de Donn\'ees astronomiques de Strasbourg, France \citep{2000AAS..143...23O}. This work presents results from the European Space Agency (ESA) space mission Gaia. Gaia data are being processed by the Gaia Data Processing and Analysis Consortium (DPAC). Funding for the DPAC is provided by national institutions, in particular the institutions participating in the Gaia MultiLateral Agreement (MLA). The Gaia mission website is https://www.cosmos.esa.int/gaia. The Gaia archive website is https://archives.esac.esa.int/gaia. J.~F. acknowledges support from NASA TESS award 80NSSC21K0792 as well as support from the Heising Simons Foundation. J.~G. acknowledges the support of the Natural Sciences and Engineering Research Council of Canada (NSERC), funding reference number RGPIN-2021-03121. This research was supported in part by the National Science Foundation under Grant No. NSF PHY-1748958. The TESS dataset is available at MAST: \dataset[10.17909/fwdt-2x66]{\doi{10.17909/fwdt-2x66}}. All the {\it AllWISE}, {\it 2MASS}, Gaia DR1/TGAS, Gaia DR2, Gaia DR3 and CatWISE preliminary data used in this paper can be found at IPAC : \dataset[10.26131/IRSA1]{\doi{10.26131/IRSA1}}, \dataset[10.26131/IRSA2]{\doi{10.26131/IRSA2}}, \dataset[10.26131/IRSA15]{\doi{10.26131/IRSA15}}, \dataset[10.26131/IRSA12]{\doi{10.26131/IRSA12}}, \dataset[10.26131/IRSA544]{\doi{10.26131/IRSA544}} and \dataset[10.26131/IRSA126]{\doi{10.26131/IRSA126}}.

\facility{Exoplanet Archive}

\pagebreak
\pagebreak
\newpage

\startlongtable
\tabletypesize{\footnotesize}
\tablewidth{0.985\textwidth}
\begin{deluxetable*}{lhhchhhhcclccc}
\tablecolumns{14}
\tablecaption{List of Members and Candidate Members of the Oceanus moving group .\label{tab:candidates}}
\tablehead{ & Gaia~DR3 & & Spectral & R.A. & Decl. & $G$ & $W2$  & & $\Delta UVW$ & & & & \\
Designation & source\_id & Sample\tablenotemark{a} & Type & (hh:mm:ss.ss) & (dd:mm:ss.s) & (mag) & (mag) & Youth\tablenotemark{b} & (\kms) & Obs.\tablenotemark{c} & Memb.\tablenotemark{d} & Rej.\tablenotemark{e} & Ref.}
\startdata
\sidehead{\textbf{Bona fide members}}
HD~77006~A & 1015799283499485440 & Gaia & G5 & 09:01:37 & 49:44:11 & 7.802 & 6.245 & uv;xr & 1.07 & $\mu\nu\varpi$ & OCEMG (97.1\%) & $\cdots$ & 4\\
\  --- HD~77006~B & 1015799283498356224 & Comp & (M3) & 09:01:37 & 49:44:09 & 12.299 & $\cdots$ & $\cdots$ & 1.09 & $\mu\varpi$ & COL (67.2\%) & $\cdots$ & 4\\
TYC~3424--1000--1 & 1011277541930006784 & HDBS & (K4) & 09:13:53 & 46:29:25 & 10.462 & 8.086 & gy;ac & 1.04 & $\mu\nu\varpi$ & OCEMG (99.8\%) & $\cdots$ & 4\\
Luhman~16~A & 5353626573555863424 & UCS & L7.5 & 10:49:14 & --53:19:05.0 & 16.945 & $\cdots$ & dm & 1.36 & $\mu\nu\varpi$ & OCEMG (90.0\%) & $\cdots$ & 8\\
\  --- Luhman~16~B & $\cdots$ & UCS & T0.5 & 10:49:14 & --53:19:05.8 & 16.962 & $\cdots$ & dm & $\cdots$ & $\cdots$ & OCEMG (90.0\%) & $\cdots$ & 8\\
LSPM~J1338+0744 & 3718565543663057152 & Gaia & M3 & 13:38:35 & 07:44:46 & 14.325 & 10.206 & uv;H$\alpha$ & 0.53 & $\mu\nu\varpi$ & OCEMG (98.4\%) & $\cdots$ & 1\\
StKM~1--1254~A & 1193907316656812800 & HDBS & K4/5 & 15:33:25 & 13:40:58 & 10.478 & 7.81 & uv & 1.41 & $\mu\nu\varpi$ & OCEMG (99.4\%) & $\cdots$ & 5\\
\ --- StKM~1--1254~B & 1193907346721779328 & Comp & (M7) & 15:33:26 & 13:40:58 & 17.039 & $\cdots$ & $\cdots$ & 1.49 & $\mu\varpi$ & OCEMG (99.7\%) & $\cdots$ & 6\\
\enddata
\end{deluxetable*}

\startlongtable
\tabletypesize{\footnotesize}
\tablewidth{0.985\textwidth}
\begin{deluxetable*}{lhhhhcccccc}
\tablecolumns{11}
\tablecaption{Signs of youth for members of the Oceanus moving group .\label{tab:youth}}
\tablehead{ & Gaia~DR3 & memtype & R.A. & Decl. & $P_{\rm rot}$ & $\log R^\prime_{\rm HK}$ & $NUV$ & $\log L_X$ & $\alpha$ index\tablenotemark{a} & \\
Designation & source id & & (hh:mm:ss.ss) & (dd:mm:ss.s) & (days) & (dex) & (mag) & (dex) & ($\times 10^{-2}$) & Refs.}
\startdata
\sidehead{\textbf{Bona fide members}}
HD~77006~A & 1015799283499485440 & BF & 09:01:37 & 49:44:11 & $\cdots$ & $\cdots$ & $13.375 \pm 0.004$ & $29.2 \pm 0.1$ & $1.04 \pm 0.01$ & -,-\\
TYC~3424--1000--1 & 1011277541930006784 & BF & 09:13:53 & 46:29:25 & 9.60142 & $\cdots$ & $\cdots$ & $\cdots$ & $3.98 \pm 0.04$ & 1,-\\
LSPM~J1338+0744 & 3718565543663057152 & BF & 13:38:35 & 07:44:46 & 1.47398 & $\cdots$ & $22.4 \pm 0.5$ & $\cdots$ & $\cdots$ & 1,-\\
StKM~1--1254~A & 1193907316656812800 & BF & 15:33:25 & 13:40:58 & $\cdots$ & $\cdots$ & $19.5 \pm 0.10$ & $\cdots$ & $1.50 \pm 0.04$ & -,-\\
HD~143295~A & 4398623381051080064 & BF & 15:59:42 & -05:04:36 & $\cdots$ & $\cdots$ & $16.27 \pm 0.02$ & $\cdots$ & $3.99 \pm 0.02$ & -,-\\
\  --- HD~143295~B & 4398623381046658176 & BF & 15:59:44 & -05:05:04 & $\cdots$ & $\cdots$ & $20.5 \pm 0.2$ & $\cdots$ & $0.60 \pm 0.08$ & -,-\\
Gaia~DR3~1426873554041125120 & 1426873554041125120 & BF & 16:31:02 & 54:13:37 & 0.74212 & $\cdots$ & $21.5 \pm 0.2$ & $\cdots$ & $\cdots$ & 1,-\\
BD+18~3497~A & 4503423641091793152 & BF & 17:55:45 & 18:30:01 & $\cdots$ & $-4.301$ & $\cdots$ & $28.0 \pm 0.2$ & $1.84 \pm 0.02$ & -,2\\
\enddata
\tablenotetext{a}{Gaia~DR3 activity index. See Section~\ref{sec:age} for more details.}
\tablerefs{References are given for rotation periods and $\log R^\prime_{\rm HK}$ indices, respectively. (1)~This paper; (2)~\citealt{2000AAS..142..275S}; (3)~\citealt{2010ApJ...725..875I}; (4)~\citealt{2018AA...616A.108B}; (5)~\citealt{2006SASS...25...47W}; (6)~\citealt{2018AJ....155...39O}; (7)~\citealt{1996ApJ...457L..99B}.}
\end{deluxetable*}

\tablewidth{0.485\textwidth}
\begin{deluxetable*}{lcccccc}
\tablecolumns{7}
\tablecaption{Radial velocities for Luhman~16~AB.\label{tab:luhmanrvs}}
\tablehead{Instrument & Date & RV$_{\rm A}$ & RV$_{\rm B}$ & RV$_{\rm AB}$ & RV$_{\rm CM}^\mathsection$ & Ref.\\
 & yymmdd & (\kms) & (\kms) & (\kms) & (\kms) & }
\startdata
    FIRE, Magellan & 130428 & $17.3 \pm 3.1$ & $11.8 \pm 3.2$ & $\cdots$ & $14.9 \pm 2.2$ & (1)\\
    CRIRES, ESO & 130505 & $20.1 \pm 0.5$ & $17.4 \pm 0.5$ & $\cdots$ & $18.9 \pm 0.4$ & (2)\\
    X-Shooter, VLT & 130611 & $25.1 \pm 7.6$ & $19.0 \pm 7.2$ & $\cdots$ & $22.4 \pm 5.3$ & (3)\\
    RSS, SALT & 130912 & $23.1 \pm 1.1$ & $19.5 \pm 1.2$ & $\cdots$ & $21.5 \pm 0.8$ & (4)\\
    FIRE, Magellan & 151220 & $\cdots$ & $\cdots$ & $13.9 \pm 3.1$ & $14.0 \pm 3.1^\dagger$ & (5)\\
    FIRE, Magellan & 160623 & $\cdots$ & $\cdots$ & $15.4 \pm 3.1$ & $15.5 \pm 3.1^\dagger$ & (5)\\
    Combined & $\cdots$ & $\cdots$ & $\cdots$ & $\cdots$ & $17.6 \pm 1.6$ & (5)\\
\enddata
\tablenotetext{}{$^\mathsection$\ The center-of-mass radial velocities ${\rm RV}_{\rm CM}$ assume a mass ratio of $0.78 \pm 0.10$ \citep{2018AA...618A.111L}.}
\tablenotetext{}{$^\dagger$\ These values assume were calculated by assuming that the A and B components contributed equally to the measured radial velocity, and uses the average relative radial velocity of ${\rm RV}_{\rm A} - {\rm RV}_{\rm B} = 2.18 \pm 0.12$\,\kms\ \citep{2018AA...618A.111L}.}
\tablerefs{(1)~\citealt{2014ApJ...790...90F}; (2)~\citealt{2014Natur.505..654C}; (3)~\citealt{2015AA...581A..73L}; (4)~\citealt{2013ApJ...770..124K}; (5)~This paper.}
\end{deluxetable*}

\startlongtable
\tabletypesize{\footnotesize}
\tablewidth{0.985\textwidth}
\begin{deluxetable*}{lhhhhccccccc}
\tablecolumns{12}
\tablecaption{Kinematics of the Oceanus moving group.\label{tab:kin}}
\tablehead{ & Gaia~DR3 & memtype & R.A. & Decl. & $\mu_\alpha\cos\delta$ & $\mu_\delta$ & RV & Num. & Plx & Gaia~DR3 & \\
Designation & source id & & (hh:mm:ss.ss) & (dd:mm:ss.s) & (\masyr) & (\masyr) & (\kms) & RVs\tablenotemark{a} & (mas) & RUWE & Refs.}
\startdata
\sidehead{\textbf{Bona fide members}}
HD~77006~A & 1015799283499485440 & BF & 09:01:37 & 49:44:11 & $-85.01 \pm 0.02$ & $-128.87 \pm 0.02$ & $4.7 \pm 0.4$ & 35 & $22.86 \pm 0.03$ & 1.1 & 1,1,1\\
\  --- HD~77006~B & 1015799283498356224 & BF & 09:01:37 & 49:44:09 & $-70.5 \pm 0.5$ & $-107.3 \pm 0.3$ & $\cdots$ & $\cdots$ & $22.9 \pm 0.2$ & 5.9 & 1,1,1\\
TYC~3424--1000--1 & 1011277541930006784 & BF & 09:13:53 & 46:29:25 & $-74.40 \pm 0.01$ & $-95.59 \pm 0.01$ & $6.9 \pm 0.2$ & 24 & $18.52 \pm 0.01$ & 1.1 & 1,1,1\\
Luhman~16~A & 5353626573555863424 & BF & 10:49:14 & -53:19:05 & $-2754.77 \pm 0.09$ & $358.72 \pm 0.09$ & $19 \pm 2$ & 6 & $500.5 \pm 0.1$ & $\cdots$ & 2,3,3\\
\  --- Luhman~16~B &  & BF &  &  &  &  &  &  &  &  & \\
LSPM~J1338+0744 & 3718565543663057152 & BF & 13:38:35 & 07:44:46 & $-145.78 \pm 0.03$ & $-63.46 \pm 0.02$ & $-9 \pm 2$ & 14 & $25.11 \pm 0.03$ & 1.2 & 1,1,1\\
StKM~1--1254~A & 1193907316656812800 & BF & 15:33:25 & 13:40:58 & $-92.39 \pm 0.02$ & $-41.79 \pm 0.01$ & $-22.3 \pm 0.2$ & 13 & $23.34 \pm 0.01$ & 1.1 & 1,1,1\\
\ --- StKM~1--1254~B & 1193907346721779328 & BF & 15:33:26 & 13:40:58 & $-93.74 \pm 0.09$ & $-44.10 \pm 0.07$ & $\cdots$ & $\cdots$ & $23.30 \pm 0.07$ & 1.0 & 1,1,1\\
\enddata
\tablecomments{The Gaia~DR3 source identifiers of all stars in this table are available in the online version.}
\tablenotetext{a}{Number of individual radial velocity measurements that were combined into the measurement reported here. See Table~\ref{tab:rvs} for individual measurements and their references.}
\tablerefs{References are given for coordinates, proper motions and parallaxes, respectively. (1)~\citealt{2021AA...649A...1G}; (2)~\citealt{2018AA...616A...1G}; (3)~\citealt{2015MNRAS.453L.103S}; (4)~\citealt{2021ApJS..253....8M}; (5)~\citealt{2020AJ....159..257B}; (6)~\citealt{2007MNRAS.379.1599L}; (7)~\citealt{2013MNRAS.433..457B}; (8)~\citealt{2016AA...595A...2G}; (9)~\citealt{2007AA...474..653V}; (10)~\citealt{2016ApJ...833...96L}.}
\end{deluxetable*}

\startlongtable
\tabletypesize{\footnotesize}
\tablewidth{0.985\textwidth}
\begin{deluxetable*}{lhhhhcccc}
\tablecolumns{9}
\tablecaption{Radial velocities of the Oceanus moving group .\label{tab:rvs}}
\tablehead{ & Gaia~DR3 & memtype & R.A. & Decl. & RV & Num. & Epoch & \\
Designation & source id & & (hh:mm:ss.ss) & (dd:mm:ss.s) & (\kms) & RVs\tablenotemark{a} & (yr) & Ref.}
\startdata
\sidehead{\textbf{Bona fide members}}
HD 77006 A & 1015799283499485440 & BF & 09:01:36.897 & 49:44:10.52 & $4.0 \pm 0.3$ & 1 & $\cdots$ & 1\\
HD 77006 A & 1015799283499485440 & BF & 09:01:36.897 & 49:44:10.52 & $3.8 \pm 0.2$ & 3 & $\cdots$ & 2\\
HD 77006 A & 1015799283499485440 & BF & 09:01:36.897 & 49:44:10.52 & $4.70 \pm 0.01$ & 1 & 2005.0753 & 3\\
HD 77006 A & 1015799283499485440 & BF & 09:01:36.897 & 49:44:10.52 & $4.847 \pm 0.002$ & 1 & 2011.0999 & 3\\
HD 77006 A & 1015799283499485440 & BF & 09:01:36.897 & 49:44:10.52 & $4.859 \pm 0.002$ & 1 & 2011.1602 & 3\\
HD 77006 A & 1015799283499485440 & BF & 09:01:36.897 & 49:44:10.52 & $4.842 \pm 0.002$ & 1 & 2011.1711 & 3\\
HD 77006 A & 1015799283499485440 & BF & 09:01:36.897 & 49:44:10.52 & $5.0 \pm 0.2$ & 16 & 2015.5 & 4\\
HD 77006 A & 1015799283499485440 & BF & 09:01:36.897 & 49:44:10.52 & $5.0 \pm 0.1$ & 28 & 2016 & 5\\
\enddata
\tablecomments{The Gaia~DR3 source identifiers of all stars in this table are available in the online version. Radial velocities for Luhman~16~AB are reported in Table~\ref{tab:luhmanrvs}.}
\tablenotetext{a}{Number of individual radial velocity measurements that were combined by the original literature reference into the measurement reported here.}
\tablerefs{(1)~\citealt{2006AAT...25..145G}; (2)~\citealt{2007AN....328..889K}; (3)~\citealt{2018AA...616A...7S}; (4)~\citealt{2018AA...616A...1G}; (5)~\citealt{2022arXiv220800211G}; (6)~\citealt{2010AA...514A..97L}; (7)~\citealt{2016AA...596A.116S}; (8)~\citealt{2020AJ....160..120J}; (9)~\citealt{2006AA...460..695T}; (10)~\citealt{2020AA...636A..74T}; (11)~\citealt{2020AA...636A..36L}; (12)~\citealt{2014AJ....147...20N}; (13)~\citealt{2002AJ....123.3356G}; (14)~\citealt{2012ApJ...758...56S}; (15)~\citealt{2010AA...521A..12M}; (16)~\citealt{2010ApJ...723..684B}; (17)~\citealt{2018ApJS..235...42A}; (18)~\citealt{2016ApJS..225...10F}.}
\end{deluxetable*}

\startlongtable
\tabletypesize{\footnotesize}
\tablewidth{0.985\textwidth}
\begin{deluxetable*}{lhhhhcccccc}
\tablecolumns{11}
\tablecaption{Galactic coordinates and space velocities of the Oceanus moving group .\label{tab:xyzuvw}}
\tablehead{ & Gaia~DR3 & memtype & R.A. & Decl. & $X$ & $Y$ & $Z$ & $U$ & $V$ & $W$ \\
Designation & source id & & (hh:mm:ss.ss) & (dd:mm:ss.s) & (pc) & (pc) & (pc) & (\kms) & (\kms) & (\kms)}
\startdata
\sidehead{\textbf{Bona fide members}}
HD~77006~A & 1015799283499485440 & BF & 09:01:37 & 49:44:11 & $-32.38 \pm 0.04$ & $6.157 \pm 0.007$ & $28.75 \pm 0.03$ & $-18.5 \pm 0.3$ & $-25.27 \pm 0.07$ & $-8.3 \pm 0.3$\\
\  --- HD~77006~B & 1015799283498356224 & BF & 09:01:37 & 49:44:09 & $-32.3 \pm 0.2$ & $6.14 \pm 0.04$ & $28.7 \pm 0.2$ & $\cdots$ & $\cdots$ & $\cdots$\\
TYC~3424--1000--1 & 1011277541930006784 & BF & 09:13:53 & 46:29:25 & $-38.94 \pm 0.03$ & $4.537 \pm 0.004$ & $37.12 \pm 0.03$ & $-19.6 \pm 0.1$ & $-23.92 \pm 0.03$ & $-7.5 \pm 0.1$\\
Luhman~16~A & 5353626573555863424 & BF & 10:49:14 & -53:19:05 & $0.5208 \pm 0.0001$ & $-1.9201 \pm 0.0004$ & $0.1841 \pm 0.0001$ & $-18.9 \pm 0.5$ & $-25 \pm 2$ & $-7.1 \pm 0.2$\\
\  --- Luhman~16~B &  & BF &  &  &  &  &  &  &  & \\
LSPM~J1338+0744 & 3718565543663057152 & BF & 13:38:35 & 07:44:46 & $13.74 \pm 0.02$ & $-6.390 \pm 0.007$ & $36.82 \pm 0.04$ & $-17.9 \pm 0.8$ & $-24.7 \pm 0.4$ & $-7 \pm 2$\\
StKM~1--1254~A & 1193907316656812800 & BF & 15:33:25 & 13:40:58 & $25.63 \pm 0.02$ & $10.151 \pm 0.006$ & $32.79 \pm 0.02$ & $-16.8 \pm 0.2$ & $-23.75 \pm 0.06$ & $-8.6 \pm 0.2$\\
\ --- StKM~1--1254~B & 1193907346721779328 & BF & 15:33:26 & 13:40:58 & $25.68 \pm 0.08$ & $10.17 \pm 0.03$ & $32.9 \pm 0.1$ & $\cdots$ & $\cdots$ & $\cdots$\\
\enddata
\tablecomments{The Gaia~DR3 source identifiers of all stars in this table are available in the online version.}
\end{deluxetable*}

\bibliographystyle{apj}
\bibliography{ads_library,other_references,Zenodo_Library}
\end{document}